\documentstyle[emulateapj,psfig,onecolfloat]{article}

\def\taueff{\overline{\tau}_{\rm eff}}
\def\etal{{\rm et~al.\ }}
\def\hmpc{\;h^{-1}{\rm Mpc}}
\def\invhmpc{\;h\;{\rm Mpc}^{-1}}

\def\hkpc{h^{-1}{\rm kpc}}
\def\kms{{\rm \;km\;s^{-1}}}

\def\kmsmpc{\kms\;{\rm Mpc}^{-1}}
\def\msun{{\rm M_{\odot}}}
\def\hmsun{h^{-1}{\rm M_{\odot}}}
\def\lya{Ly$\alpha$}

\def\df{\Delta^{2}_{\rm F}(k)}
\def\du{\delta_{u}}
\def\degr{^{\circ}}
\def\simlt{\lower.5ex\hbox{$\; \buildrel < \over \sim \;$}}
\def\simgt{\lower.5ex\hbox{$\; \buildrel > \over \sim \;$}}

\newcommand{\PSbox}[3]{\mbox{\rule{0in}{#3}\includegraphics{#1}\hspace{#2}}}

\begin{document}

\twocolumn[

\title{
Ionizing radiation fluctuations and large-scale structure in the
Lyman-alpha forest
}

\author{
Rupert A.C. Croft$^{1}$
}
\affil{Department of Physics, Carnegie Mellon University,
Pittsburgh, PA 15213}

\begin{abstract}
We investigate the large-scale inhomogeneities of the hydrogen
ionizing radiation
field in the Universe at redshift $z=3$.
Using a raytracing algorithm, we
simulate a model in  which quasars are the dominant sources 
of radiation. We make use of large scale 
N-body simulations of a $\Lambda$CDM universe, and include such
effects as finite quasar lifetimes and
output on the lightcone, which affects the shape
of quasar light echoes. We create \lya\ forest
spectra that would be generated in the presence of such a fluctuating
radiation field, finding that the power spectrum of the \lya\
forest can be suppressed by as much as $15\%$ for modes with 
 $k=0.05-1 \hmpc$. This relatively small effect may have 
consequences for high precision measurements of the \lya\ power 
spectrum on larger scales than have yet been published. 
We also investigate a second probe of the ionizing radiation fluctuations, 
the cross-correlation of quasar positions and the \lya\ forest.
For both quasar lifetimes which we simulate ($10^{7}$ yr and $10{^8}$yr),
we expect to see a strong decrease in the \lya\ absorption
close to other quasars (the ``foreground'' proximity effect). 
We then use data from the Sloan Digital Sky Survey First Data Release
to make an observational determination of this statistic. We find no sign of 
our predicted lack of absorption, but instead increased absorption close
to quasars. If the bursts of 
radiation from  quasars last on average $< 10^{6}$ yr, then we would
not expect to  be able to see the foreground effect. However,
the strength of the absorption itself seems to be indicative of rare objects,
and hence much longer total times of emission per
quasar. Variability of quasars in  bursts 
with timescales $> 10^{4}$yr and $<10^{6}$ yr could reconcile
these two facts.

\end{abstract}
 
\keywords{Cosmology: observations -- large-scale structure of Universe}
]

\footnotetext[1]{rcroft@cmu.edu}

\section{Introduction}
The \lya\ forest is a useful probe of the structure of the high redshift
Universe. Over most of the volume of space, the hydrogen responsible
for \lya\ absorption is in photoionization equilibrium with
a background radiation field. The optical depth for \lya\ absorption 
at a given point in space is related simply to the 
density (see e.g., Bi 1993, Hui, Gnedin \& Zhang 1997,
Croft \etal 1997) and also inversely to the 
intensity of the ionizing radiation. The correlation of \lya\
absorption with the density field has been much studied, including its
use as a probe of matter clustering (e.g., Hui 1999,
Nusser \& Haehnelt 1999, McDonald \etal 2000, Viel \etal 2002).
 The inhomogeneities of the
radiation field however have not been much examined in this context
(although see the recent work of Meiksin \& White 2003ab). In this
paper we present a method for predicting the large-scale fluctuations
in the radiation field and their effect on \lya\ forest spectra
and their statistical properties.

The study of the reionization of the Universe using radiative transfer
in simulations is a rapidly
growing field (e.g., Abel \etal 1999,
 Gnedin 2000,  Ciardi et al 2001,
Sokasian \etal 2001,  Razoumov \etal 1999,2002)
At redshifts soon after the reionization epoch, the Universe is 
still relatively optically thick, and the fluctuations in the 
radiation field are expected
to be very large (e.g., Meiksin \& White 2003b).
 As the Universe expands, however, the dilution
of the density field reduces the number density of neutral hydrogen atoms
also, so that by $z=3$ (the epoch we will study in this paper), the
mean free path of ionizing photons is expected to be $\simgt 100 \hmpc$ 
comoving
(Haardt and Madau 1996, hereafter HM96). If the dominant sources of 
photons are rare objects such as quasars, then only a few will
lie within each attentuation radius (the mean distance
to reach a unit optical depth for absorption). Fluctuations in the
radiation field will be relatively gentle, but occur on large scales,
comparable to the mean separation between sources (e.g., Zuo 1992a,
Fardal and Shull 1993).
This epoch is difficult to treat accurately with simulations, because one
would like to resolve both the small scale clumping of the IGM and the 
large distance between rare sources. In this paper we will make use 
of hybrid approach which combines large dark matter simulations with 
optical depths calibrated from a high resolution hydrodynamic run.

The intensity 
 fluctuations in the ionizing background have been 
studied using an analytical technique by Zuo (1992a,b) for randomly
distributed sources. The effect on the \lya\ forest 
was also studied by Fardal and Shull (1993) who used the same techniques,
as well as Monte Carlo simulations, again for randomly distributed clouds.
Croft \etal (1999) made a simple study of the effect of such fluctuations 
on the recovery of the matter power spectrum from the \lya\ forest,
finding no significant effect on the small scales 
($k > 0.2 \invhmpc$) then observationally
accessible but a potentially interesting effect on large scales.
All of these approaches assumed a uniform IGM which attentuates 
photons isotropically (no shadowing is possible), as well
as ignoring the effect of finite source lifetimes. More recent
work has been carried out by Gnedin and Hamilton
(2002), in a small simulation volume ($4 \hmpc$) at $z=4$, finding that 
fluctuations are negligible on these scales and below. Meiksin and White
(2003ab) find strong fluctuations at $z>5$, combining a PM dark matter
simulation with a uniform attentuation approximation.

The redshift ($z=3$) which we focus on in this paper is one for 
which much observational data is available, both for the \lya\
forest, and for Lyman Break Galaxies (e.g., Adelberger \etal 2003, 
Steidel \etal 2003 ).
The structure in the radiation field itself is expected to be quite interesting
and complex, manifesting itself on much larger scales than the
structures in the density field. 
We aim to investigate how the clustering of the \lya\ forest 
will be influenced by the structure of the radiation field,
which in turn depends on  the lifetime of quasar sources, and the 
fractional contribution of quasars to the overall background radiation 
intensity. We will also investigate the \lya\ forest 
absorption in lines of sight which pass close to foreground 
quasars (testing what is 
sometimes known as the foreground proximity effect). Here,
close to the sources of radiation,
the effect of inhomogeneities in the radiation field should
manifest themselves most strongly. Comparison with observational
measurements have the potential to constrain the source lifetime and possibly
the geometry of space.

The paper is set out as follows. In \S2, we describe the set of 
simulations which we will use. In \S3 we describe our raytracing method
which we we apply to the outputs of these simulations. We detail some
properties of the resulting radiation field before describing our
procedure for generating \lya\ forest spectra in the presence of
this field. In \S4, we measure the power spectum of the
\lya\ forest 
flux in these spectra, and investigate its dependence on quasar lifetime,
beaming of radiation, and the inclusion of light cone effects.
In \S5 we turn to the \lya\ forest averaged around foreground quasars,
showing simulation results. We also compute this statistic from
the Sloan Digital Sky Survey First Data Release and carry out a 
comparison with the models. Our summary and discussion form \S6.

\section{Simulations}

We will use N-body simulations of large-scale structure in order to
create a framework involving sources and sinks of radiation.
Our aim is to examine the statistical properties of the radiation
field on large scales, and to make use of sparsely distributed sources
(QSOs). Because of both these factors, we must use large simulation volumes,
and we choose to run several realizations with different random phases.
We also have run different box sizes with different mass resolutions in order
to carry out a resolution study. The main
set of simulations which we will use are dark matter only. In order to
compute the optical depth to absorption for ionizing photons
which pass through the simulation, we have adopted a hybrid approach,
making use of an output of a full hydrodynamic simulation.
 We calibrate the optical depths associated with the 
dark matter in our large low resolution simulations using information
from the high resolution hydro run. In the near future, when
large box high resolution gas dynamical simulations become available,
this hybrid approach will not be necessary.

The cosmological model we choose to simulate is a 
cosmological constant-dominated CDM universe, consistent with 
measurements from the WMAP satellite (Bennett \etal 2003, Spergel \etal 2003).
We use with 
$\Omega_{\Lambda}=0.7$, $\Omega_{\rm m}=0.3$ $\Omega_{\rm b}=0.04$,
and a Hubble constant $H_{0}=67 \kmsmpc$. The initial linear power
spectrum is cluster-normalized with a linearly extrapolated amplitude
of $\sigma_{8}=0.9$ at $z=0$. 

\subsection{Collisionless dark matter only}

As stated above, we use dark matter only simulations for most of
the work in this
paper, which were run with the $P^{3}M$ code of Efstathiou \etal (1985).
We have three sets of simulations, each with $200^{3}$ particles,
and within each set run 5 realizations with different random phases.
Unless stated otherwise, our results will be averages over the 5 realizations
in each case. Our three sets of simulations have different box sizes
and force resolutions. The box sizes are $500 \hmpc$, $250 \hmpc$ and
$125 \hmpc$ for what we term the ``d500'', ``d250'' and ``d125'' 
sets. Their comoving force resolutions are $130 \hkpc$, $65 \hkpc$ and 
$33 \hkpc$ respectively, and they were all run from redshift $z=24$ to 
redshift $z=3$. The mass per particle for the runs is $1.3 \times 10^{12}
h^{-1} \msun$ for the d500, $1.6 \times 10^{11} h^{-1} \msun$ for d250,
 and $2.0 \times 10^{10}h^{-1} \msun$ for d125.

\subsection{Hydrodynamic simulation}
We make use of the $z=3$ simulation output of the simulation
described in White, Hernquist \& Springel (2001), and Croft \etal (2002),
and which was kindly provided by Volker Springel and Lars Hernquist.
The box size is $33.5 \hmpc$, and there are $300^3$
particles representing the gas, and $300^3$ representing the
collisionless dark matter. The particle masses are therefore
$1.5 \times 10^{7}\, \hmsun$ for the baryons and $1\times10^{8}\,
\hmsun$ for the DM. The simulation was run using the smoothed particle
hydrodynamics (SPH) code {\small GADGET} (Springel \etal 2001),
and includes gas dynamics, cooling, and a multiphase treatment
of star formation (see Springel \& Hernquist 2003). Also,
importantly, it was run including a uniform background of ionizing 
radiation, based on the results of HM96.

\section{Method}

Solving the general radiative transfer (RT) problem in the Universe
is extremely challenging. Depending on the specific application,
there are many shortcuts and innovative treatments which can be used.
For example, when looking at growth of ionized regions in an
optically thick neutral medium, the jump condition can be 
solved by casting along rays (Sokasian, Abel and Hernquist, 2001). 
A summary of the many different recent approaches to solving RT
in the early Universe can be found in Section 1 of Maselli, Ferrara and
Ciardi (2003).

In this paper, we are interested in the Universe at $z=3$, which
is close to being optically thin, 
the mean free path of hydrogen ionizing
photons being $\sim 100 \hmpc$. The simplest 
approximation which can be made is that the attenuation of
photons is isotropic about each source (e.g., Fardal \& Shull 1993, 
Meiksin \& White 2003). Once this assumption is dropped, it becomes 
necessary to deal with inhomogeneities in the optical depth, 
which we do using a raytracing approach. We make many other simplifications,
however. For example, we do not carry out time dependent raytracing,
so that the effect of earlier passages of radiation has no effect
on the propagation of photons passing through the same region of space.
As the fluctuations in the radiation field are small, and the time
taken to recover to the mean neutral fraction is short, this is not 
unreasonable (we shall return to this later). Also, for simplicity,
we deal with only one frequency, photons with energy 1 Rydberg.
 
\subsection{Raytracing}

Starting from sources of radiation (we describe their selection 
in \S 3.4 below), we trace the paths of photons through the static
density field. We adopt a Monte Carlo approach (see e.g., Ciardi \etal 1999,
Maselli \etal 2003), in which photons are emitted isotropically 
from sources (this is relaxed later for beamed sources). The
directional polar angles are randomly picked from the distributions:
\begin{equation}
\theta=\arccos(1-2R_{\theta}),\ \phi=2\pi R_{\phi},
\end{equation}
where $R_{\theta}$ and $R_{\phi}$ are random numbers from 0 to 1 
(Ciardi \etal 1999).

Our approach during raytracing is not to first assign particle
positions and densities to a grid, but instead integrate through the
actual distribution of particle kernels. Our two reasons for doing this are to
allow for more accurate shadowing from matter close to a source, and 
to be closer to the spirit of the Lagrangian technique involved.
 Since we are using dark matter particles as
a proxy for gas, we first compute the optical
depth for absorption for a given mass and density of 
dark matter. We calibrate the relation using the hydrodynamic simulation
as follows. We first decide
on a dark matter mass, which we take to be 8 of the 
dark matter only simulation particles.  We then randomly pick cubes which
contain this mass of dark matter plus gas from the hydrodynamic 
simulation, picking the
centers at random and varying the side length so that they contain the
required mass. We then shoot rays through the cube and for each one
compute the optical depth for absorption by the neutral hydrogen
present:
\begin{equation}
\tau(\nu)=\sigma(\nu)\int n_{HI}(x) dx,
\end{equation}
where $\sigma(\nu)$ is the photoionization cross section.
In this way, we make use of the high resolution of the hydrodynamic
simulation which allows for clumping in the gas. Averaging over
a number of sightlines ($N$) gives us an effective optical depth
due to the cube with density $\rho$:
\begin{equation}
\tau_{\rm eff}(\rho)=-\ln \left(\frac{1}{N} \sum_{i=1}^{N} e^{-\tau_{i}}\right)
\end{equation}

\begin{figure}[t]
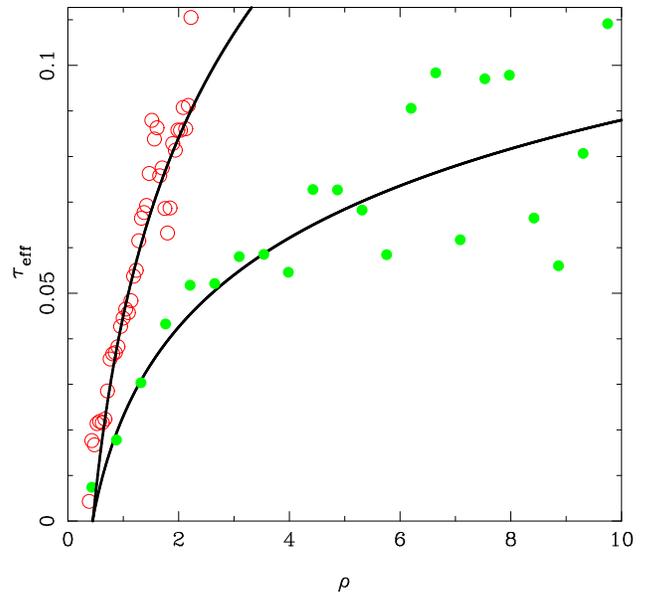

\centering
\PSbox{taueffvsrho.ps angle=-90 voffset=265 hoffset=-50 vscale=46 hscale=46}
{3.5in}{3.3in} 
\caption[tauvsrho]{ 
\label{tauvsrho}
The effective optical depth $\taueff$ seen by photon crossing through
the neutral hydrogen associated with dark matter with the mass of 8 
dark matter simulation particles, as a function
of density in units of the mean. We show results from the
d250 simulation (green points) and d500 simulation (red points),
as well as parametric fits (solid lines). 
 }
\end{figure}

In Figure \ref{tauvsrho}, we show $\taueff(\rho)$ as a function of the
density of matter in the cube, $\rho$. We show results for two dark matter
masses, appropriate for 8 particles from 
 the d500 and the d250 simulations. In the case
of the d500 runs, the side length of a cube with the mean density 
that contains 8 particles is $5 \hmpc$ (comoving).
 From  Figure \ref{tauvsrho}, we
can see that at the mean density, $\taueff$ is $0.045$. This means
that to reach an optical depth of 1
a photon with energy 1 Ryd.
would have to move through a universe (uniform in density
on scales $>5 \hmpc$) a distance of $\sim5 \hmpc \times 1/0.045$,
 or $111 \hmpc$.

This figure can be compared to the attenuation length $r_0$, which
has been computed by many authors (e.g., Paresce \etal  1980)
from  the assumption of a universe uniform on large scales but populated
with \lya\ clouds and Lyman limit systems with different column
densities, specified by matching with observations. Depending on the
how the observations are interpreted and which cosmology is used, values of 
$r_0$ at $z=3$ (converted to comoving units) including 128 $\hmpc$ (HM96),
$100-375 \hmpc$ (Fardal and Shull 93) have been previously found. In Meiksin
and White (2003a,b), $r_0$ was computed from dark matter based simulations 
of the \lya\ forest with the addition of observational data on
Lyman-limit systems which were unresolved by the simulation. Meiksin \& 
White (2003a,b) find
$r_{\rm att}\sim 210 \hmpc$ at $z=3$.
 Our simulation result is therefore on the 
low end compared to other computed estimates, something which will
tend to maximize the fluctuations in the radiation field. We note, however
that because in our case we deal with an inhomogeneous absorbing medium,
the effective attentuation length seen by photons from different sources
and travelling in different directions will vary, and that the value of
$r_0$ is not directly comparable. For example, in our $500 \hmpc$
box runs (for QSOs with lifetimes $10^{7}$ yr - see later), 
we find that $30\%$ of 
photons are not absorbed after travelling half a box length, and those that
are absorbed travel a mean distance of $80 \hmpc$, so that the effective
value of $r_{\rm att}$ is greater than $0.3\times 250+ 0.7\times80= 130 \hmpc$.

In  Figure \ref{tauvsrho} we show a simple functional form
which reproduces the trend of the $\rho-\taueff$
 datapoints adequately, and which 
we use to calibrate our dark matter simulations:
\begin{equation}
\tau_{\rm eff}(\rho)=\frac{b}{500 \hmpc}(0.13 \log_{10}
\rho+0.045), \label{taueff}
\end{equation}
where $b$ is the simulation box size in $\hmpc$.
If $\taueff < 0$ we set $\taueff=0$.
We use this relation to carry out our raytracing. Each photon
is tracked through the simulation until it has passed through an
accumulated optical depth of 
\begin{equation}
\tau > -\ln(1-R)
\end{equation}
where $R$ is a random number uniformly distributed
 between 0 and 1 (Ciardi \etal 1999), 
or if it has travelled more than one half the side length of the simulation
box. 

In the ideal case where we would be using a large hydrodynamic simulation
to represent the density field and associated neutral hydrogen, we would
integrate through the particles' SPH kernels (see e.g., 
Kessel-Deynet \& Burkert, 2000 for an SPH-based approach to RT).
 In the present
case, we have dark matter particles, and we treat each as if it where
distributed in space as a spherical top hat with radius $r_{h}$ given
by the distance to the 8th nearest neighbour particle.
Each dark matter particle then has a characteristic density
given by the particle mass divided by the volume of this sphere.
  We then assign
(1/8) $\taueff$ (Equation \ref{taueff})
 to the $\tau$ sum of any photon path which comes within $r_{h}$ of the
particle center. At the same time, we keep track of the
radiation intensity in terms of the number
of photons per unit area which have passed through the particle.
This way each photon imparts information to on average $\sim 100$
particles before it is absorbed or has travelled more than 
half the box length.

As stated before, when photons are absorbed in a particle, we do not 
update any ionization states, as we are making the simplification
of a time independent treatment. We will instead use the calculated
radiation intensities together with the assumption of photoionization
equilibrium when we calculate our \lya\ spectra. Although the 
recombination time for gas with the low densities we are
considering here is long ($\simgt 1$ Gyr), the relevant timescale
is the equilibration time. This is the time that the highly ionized gas
takes to respond to small changes in the ionizing radiation field (for example
a factor of 2 or 10) before it comes back into photoionization
equilibrium. This is closer to $\sim 10^{4}$ yrs,
much shorter than the quasar lifetimes that we will consider.
A further approximation which we make is to assume that space
is Euclidean. This is reasonable as long as the mean free path of photons
is much less than the Hubble length, which is true at these high redshifts.

As well as computing the intensity of the radiation field using raytracing
from QSO sources, we allow the addition of a uniform background field to 
account for photons for which the sources 
are more homogeneously distributed and which 
we do not attempt to simulate directly. This will include recombination 
radiation which HM96 showed contributes $\sim 30\%$ to the photoionization 
rate at $z=3$. Radiation from stars in galaxies such as Lyman-break objects
could also make up a substantial contribution to the intensity (see e.g.,
Steidel \etal 2001, Haenhelt \etal 2001, Sokasian \etal 2003). 
As their space
density is much higher than that of QSOs, modelling them as a uniform
contribution to the ionizing radiation field at $z=3$ 
is a good approximation (Kovner and Rees 1989, Croft \etal 2002).

As far as the numerical techniques are concerned, we use a chaining
mesh (Hockney \& Eastwood 1981) to find particles whose kernels intersect
the paths of photons. Each of our fiducial simulations was run
with $32\times10^{6}$ photon packets (see convergence tests below). This 
takes $\sim 4$ hours running on 8 Pentium III 1 Ghz processors in parallel.

\subsection{Raytracing examples}

\begin{figure}[t]
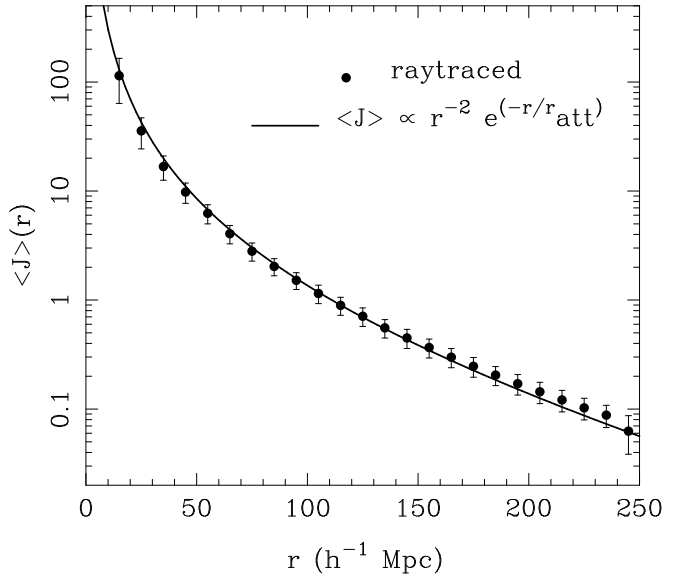

\centering
\PSbox{uniform.ps angle=-90 voffset=255 hoffset=-50 vscale=46 hscale=46}
{3.5in}{3.2in} 
\caption[uniform]{ 
A test of the raytracing method in a model of a single quasar embedded in a 
uniform medium. We plot the mean radiation intensity at a given
distance $r$ from the quasar. The solid line shows the theory curve, assuming
that the radiation attenuation length at 912 \AA\ is 
$r_{\rm att}=111 \hmpc$ 
 (measured from gasdynamical simulations). The points show the 
radiation intensity 
measured in volume filled with poisson distributed particles. A single
quasar was placed in the center of the volume and the raytracing algorithm
(\S 3.1) used to calculate the intensities. We
average the intensity in cubical cells, and the error bars on the points are 
the standard deviation among cells which fall in the same $r$ bin.

\label{uniform}
}
\end{figure}

A simple test of our raytracing scheme is to compare it with results
from the analytical solution for a single source in a uniform medium.
We set up a Poisson distributed distribution of particles in a $500 \hmpc$
box, with the same mean 
space density as in our dark matter simulations. We then
place a single source at the center, and carry out raytracing as 
detailed in Section 3.1. Each particle then has an associated 
radiation intensity
value.
 We bin the particles into $256^{3}$ cubic cells
and calculate the mean particle weighted radiation intensity for each cell.
As the density field is uniform, our estimate of $r_{\rm att}=111 \hmpc$ 
from Section 3.1) can be used for an analytical comparison.
 The radiation intensity
at a distance $r$ should therefore be 
$\propto \frac{1}{r^{2}} e^{-r/r_{\rm att}}$,
which we plot as a line in Figure \ref{uniform}. Our simulation results
are shown as points, with error bars representing the standard
deviation among cells which fall in the same $r$ bin. The Monte Carlo
raytraced photons reproduce the analytical curve reasonably well,
indicating that the numerical techniques appear to be working 
satisfactorily. As further tests of our techniques, this time for  
non-uniform media, we carry out resolution tests (later in the
paper, \S4.1 and \S5.2)
making use of our suite of simulations with different box sizes.

\begin{figure}[t]
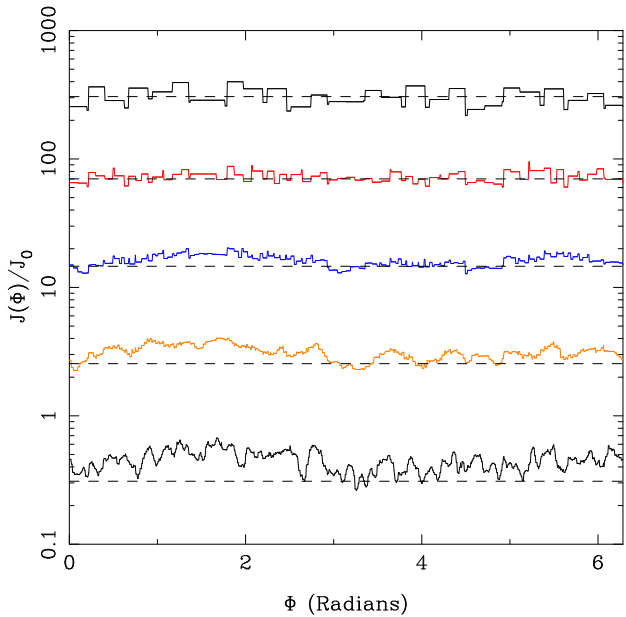

\centering
\PSbox{ringtest.ps angle=-90 voffset=260 hoffset=-45 vscale=46 hscale=46}
{3.5in}{3.2in} 
\caption[ringtest]{ 
The intensity as a function of angle in rings at a distance of $10, 20, 40, 80,
160 \hmpc$ from a single quasar in the simulation density field.
The straight lines show the uniform attentuation approximation.
Normalized so that $J=1$ at $r=r_{\rm att}$ for uniform approximation.
\label{ringtest}
}
\end{figure}

It is interesting to look at the deviations from the uniform attentuation
approximation, again for a single source, in order to see if the
effect of reduced absorption expected in voids, and enhanced absorption
close to sources in dense regions do lead to additional fluctuations
in the radiation intensity. We have again placed a single source in
the center of our computational volume, but this time we use
the density field from one of the d500 simulations. The source was chosen
to be at the position of one of the bright quasars (see \S 3.4 below) and
then raytraced photons from it were followed through the simulation
volume. We again averaged the radiation intensity in cells of size $500/256
\hmpc$, and plot the results as a function of angle for cells in rings
(all in the same plane) at 
several different distances from  the source (Figure \ref{ringtest}).
We also show the results from the uniform attentuation approximation, 
which gives a constant intensity as a function of angle (dashed lines). 

The first thing that we  notice is that the intensity decreases with
distance, as expected. At a distance of 10 $\hmpc$, there are fluctuations
present in the simulation results. These are largely due to the fact that
as with the previous test we have rebinned the particles into cubic cells (a
$256^3$ grid) and are plotting the mean intensity in each grid cell. As
we move round the ring, the center of the closest cell to 10 $\hmpc$
can be up to $\sim 1 \hmpc$ from it, which 
leads to expected fluctuations of order $\pm 20\%$. These fluctuations
are solely an artefact of the binning used for the plot, and as expected
diminish with distance. 

As we reach the $40 \hmpc$ ring, we see clear 
variations due to the anisotropy of the optical depths. At an 
angle $\Phi \sim 1.5\ {\rm Rad}$, there is an intensity $\sim 50 \%$ higher 
than at 
$\Phi \sim 3\ {\rm Rad}$. The photons appear to have been travelling
through a void and an overdense region respectively. The differences 
between the angles increase as we move farther out. We can also see
that angles which were ``in shadow'' at small distances from the source keep
the imprint at larger distances, as expected. The angular scale of
the newly imprinted intensity fluctuations also appears to go down 
with distance, as should happen if they are associated with features
with a fixed physical scale. At a distance of $160 \hmpc$, the intensity
fluctuations can be a factor of $\sim 2$. It can also be seen
that at this distance for most of space there is a higher intensity 
than would be expected in the uniform attentuation approximation
with $r_{\rm att}=111 \hmpc$). There are however a few shadowed regions 
where photons had presumably not been moving through voids, and
which are noticeably darker.

The situation with more sources present would of course be even more
complex. One could argue that the angular variations would cancel
out as each point in space can usually see more than one source
at any given time. However, with sparse QSO sources, this number would
not be large, and these sources would have their own shadows in front
of them. We also note that it is likely that the $z=3$ situation,
with the mean  free path of photons being very large 
will be less sensitive to the difference between raytracing and the 
uniform attenuation approximation than higher redshifts. In order to
test the difference between these two methods at $z=3$, we have carried out 
simulations using the uniform attentuation approximation and
will present results for these later.

\subsection{Lightcone effects}

The light travel time across our $500 \hmpc$ box is $6 \times 10^{8}$ yrs
at $z=3$. If the source lifetime is much less than this, then we may have to 
take this into account. For example, light will take 20 Myr to 
each a point $2\times10^{7}$ light years away in space (we are using the 
Euclidean approximation). If the source has switched off by this time, then
the radiation will remain as a light echo propagating through space
(analogous to supernova light echos e.g., Crotts 1988).
This will affect the ionization state of the IGM  and the 
\lya\ forest. If we were able to recieve information from all points
in space at the same time, the light echos would appear as spherical 
shells centered on the position of each source, with a thickness equal to the
source lifetime times the speed of light.  

However, because we recieve information from the 
Universe at different times, the light echoes will not quite have this shape.
Information from points behind a source will be from an earlier
time than information that comes to us from in front of a source.
Because of this the effect of the light echoes to an observer
at a given time will not be spherically symmetric 
about the  position of the source. We model these light 
cone effects as well as the finite lifetime of sources.
In order to do this, we
to calculate for each ray that we cast a minimum and maximum distance 
along it that the effect of its radiation will be present.

If these distances $r_{min}$ and
$r_{max}$ are  in comoving $\hmpc$ then we have the following for
each source and ray:
\begin{equation}
r_{min}=A (1+z) 
\frac{c t_{off}}{1+{\hat{\underline y}}\cdot {\hat{\underline r}}}
\end{equation}
and
\begin{equation}
r_{max}=A (1+z) \frac{c t_{on}}{1+{\hat{\underline y}}\cdot
{\hat{\underline r}}}.
\end{equation}
Here $t_{on}$ and $t_{off}$ are the times that the source switched on and off,
respectively, in Myr, and $A=h/3.26$. The sources
which are deemed to be on when the simulation is ``observed'' are on
at the time the lightcone passes them.  The $t=0$ point is set for each 
for each QSO to be the time of lightcone passage.
The vector ${\hat{\underline r}}$ is the unit vector along the ray and
${\hat{\underline y}}$ is the unit vector along the line of sight direction
(chosen here to be the $y$-axis).

We note here that we do not output the evolving density field on the lightcone,
as we only make use of simulation density outputs at $z=3$. A more
sophisticated  treatment could do so (which would also mean sacrificing the
periodic boundary conditions of the simulation volume). As long as 
we are only interested in predictions for one particular redshift,
our approach is sufficient. Also, we will present results with and without
radiation lightcone effects.

\subsection{Quasar sources}
We select the positions of luminous quasars in the dark matter simulations
to be peaks in the  density field. We then assign a luminosity
to these peaks in order to reproduce the observed quasar luminosity
function. Because the dark matter simulations 
have low mass resolution
(particle masses for the
d250 series are $1.6 \times 10^{11} h^{-1}\msun$),
 we use this approach rather than 
trying to identify halos, and we do not carry out a more sophisticated 
modelling of the quasar population (see e.g., Di Matteo et al 2003 
for more detailed modelling of quasars in simulations). 

In order to select peaks in the dark matter density field, we choose not
to use a grid, but instead assign a density to each particle based
on the distance to the 8th nearest neighbor. We then loop through the
list of particles a second time, and if any particle has a higher
density that its 8 nearest neighbors, then we define it to be a peak particle,
with a peak height given by its density. 

Each quasar is assigned to a peak particle. Varying the lifetime of
 quasar sources
has potentially interesting effects, so that in this paper we will
consider two different cases, source lifetime $t_{q}=10^{7}$ years,
and $t_{q}=10^{8}$ years, chosen to bracket the expected range
derived from other observations (e.g., Steidel \etal 2002). In each case,
all quasars are chosen to have the same lifetime. We then compute the
expected space density of quasars above a given magnitude limit using
observational data (see below). From this, we compute
number of quasars in the simulation active at one
particular instant at $z=3$, $n_{now}$.  
We make the assumption that the number of quasars which 
were on at any time in the past is $n_{q}=n_{now}\times(t_{H}/t_{q})$, where
$t_{H}$ is the Hubble time at $z=3$. 
The redshift difference between the edges of the
box is $\delta_{z}=\pm0.4$, so that our assumption effectively
results in a non-evolving quasar population over that interval.
We then select the highest $n_{q}$ peaks in the box and assign to 
each one a time that it switched on between 0 and $t_{H}$. Only the
quasars which have $t_{on}<$ the light travel time across half the box
are kept. In this model,  each peak  can have only host one active 
quasar phase during the time up to $z=3$.

Each quasar is then given a luminosity by drawing randomly from the
B-band luminosity function used by HM96 and Sokasian \etal (2002):
\begin{equation}
\phi(L,z)=\frac{\phi_{*}}{L_{*}(z)}\left(\left[\frac{L}{L_{*}(z)}
\right]^{\beta_{1}}+\left[\frac{L}{L_{*}(z)}\right]^{\beta_{2}}\right)^{-1}
\end{equation}
Here $L_{*}(z)$ evolves with redshift so that:
\begin{equation}
L_{*}(z)=L_{*}(0)(1+z)^{\alpha-1}\exp [-z(z-2z_{*})/2\sigma^{2}_{*}]
\end{equation}
The luminosity function parameters are $\beta_{1}=1.83, \beta_{2}=3.7,
z_{*}=2.77, \sigma_{*}=0.91$ and $\log(\phi_{*}/{\rm Gpc}^{-3})=2.37$.
This fitting formula (see Boyle, Shanks and Peterson, 1988, Pei 1995),
) is valid for an open universe with $h=0.5$, and 
as  Sokasian \etal (2002)
 have done, we rescale the volume element and luminosity
to that appropriate for our $\Lambda$CDM cosmology. Because we only directly 
simulate quasars with luminosity greater than $M_{b}=-25$,
a portion of the background intensity may be contributed by fainter quasars.
This will happen if the luminosity function above can be 
extrapolated to fainter magnitudes than observed, giving for example
a $30\%$ higher intensity if we instead use $M_{b}<-22.5$.

We convert from a $B$-band luminosity to a source luminosity per Hz at 912\AA\ 
assuming a quasar spectral index of $\nu=-1$.
Below we will describe the intensity at $912$ \AA\ that arises from 
the sources we do simulate and how we add an additional uniform radiation
field (which we vary widely) to model other contributions (including
those of fainter quasars).

\subsubsection{Anisotropic radiation}
The ionizing continuum radiation from quasars may not be emitted 
isotropically. Models of AGN unification (see e.g., the review by Urry and
Padovani 1995) predict that an obscuring torus
of material cuts down the emitted radiation to cones of opening angle
$\sim 90\degr$. We allow for this possibility by running
models with and without this beaming of radiation (we use
``beaming'' in the rest of the paper as shorthand for this 
anisotropic emission of radiation, which should not be
confused with relativistic beaming). In the case with 
beaming, we use an opening angle of $90\degr$ and pick
the orientation of the central axis randomly. This will mean that only $29\%$
of quasars are actually visible to an 
observer at one point. In order to match
the observed number density, the actual space density of objects must therefore
be increased.

\subsubsection{Quasar clustering}

A useful check of our simulated quasar sources 
is to compute their two-point correlation function. Because more massive
host galaxies (or in our case higher peaks in the density field) 
are predicted to cluster more strongly (e.g., Bardeen \etal 1986),
the correlation function has been proposed as a potential diagnostic
of the quasar lifetime (longer lived quasars must be rarer and are
therefore likely to be in more massive galaxies) by Martini and 
Weinberg (2001), and Haiman and Hui (2001). In our case, rough consistency
with observed clustering is all we seek. 

\begin{figure}[t]
\centering
\psfig{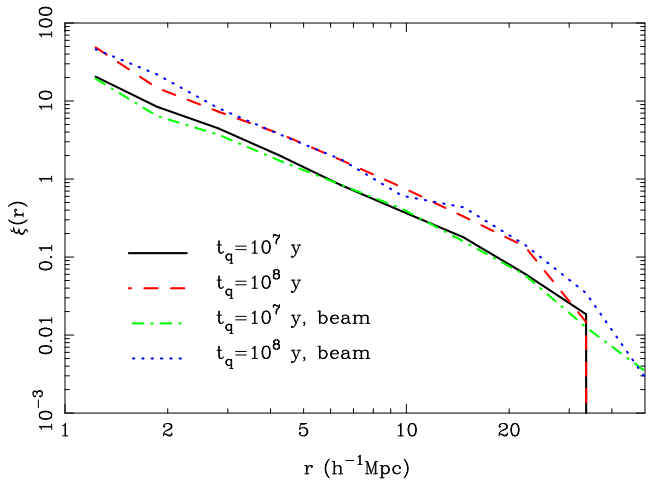}
\caption[qsoxi]{ 
\label{qsoxi} Clustering of quasars measured in real space.
We show the correlation function for two different quasar lifetimes 
and for isotropic and anisotropic emission.
}
\end{figure}

In Figure \ref{qsoxi} we show the correlation function $\xi(r)$ for quasars
with two different lifetimes $t_{q}$ and with and without beaming.
The quasars with $t_{q}=10^{7}$ yrs have a clustering
amplitude $r_{0}=6.0 \hmpc$ where $\xi(r)=(r/r_{0})^{\gamma}$ is 
the usual power
law fit. The slope $\gamma=-2.0$, and is the same for the longer lived
quasars, which have $r_{0}=8.5 \hmpc$. There is no significant
difference between the unbeamed and beamed quasars, although we might expect
the beamed quasars to have a slightly 
lower clustering amplitude because of their
higher space density. For comparison, $r_{0}=6.4 ^{+0.45}_{-0.44}$ for
quasars from the 2dF QSO survey (Croom \etal 2002). This measurement
is not however at the same redshift as our simulation, but for quasars
chosen from the
interval $z=0.3-2.9$, with a mean quasar redshift $\overline{z}=1.5$.
The slope of the 2dF correlation function is also somewhat shallower
$\gamma=-1.56$. If the clustering of observed quasars does not evolve
from $z=3$ to $z=1.5$, this means that in the context of our simulations
a lifetime close to $10^{7}$ years is favored. 
The 2dF quasars are however also fainter (by 2 absolute magnitudes in the mean)
 than our simulated
sample (which more closely matches that of the Sloan Digital Sky Survey).
This means that the 2dF quasars probably have lifetimes somewhat longer
than  $10^{7}$ years.
The modelling uncertainties
are considerable and our approach is fairly crude,
so that this analysis can only give us a rough constraint on $t_{q}$.

\subsection{Uniform background}

In addition to quasars there are at least two other  sources of 
ionizing radiation. One is recombination radiation, which HM96 compute
should account for $30\%$ of the ionizing intensity at $z=3$. The
other is galactic stars. The discovery of flux in the spectrum
of Lyman Break Galaxies (LBGs) beyond the Lyman limit (Steidel \etal 2001)
has led to some speculation that LBGs may contribute a substantial amount
 to the intensity (see e.g.,
Haenhelt \etal 2001, Hui, Zaldarriaga \& Alexander 2002).
 Sokasian, Abel and Hernquist 2003,
using radiative transfer calculations and simulations of structure
formation find that in order to match the observed opacities
of HeII and HI, the stellar contribution should contribute about 
equally with quasars at $z=3$.

Both of these sources of radiation will be distributed much more
smoothly than quasars, having a significantly higher space density. The 
fluctuations at $z=3$ resulting from them will therefore be negligible.
The radiation intensity flucutations resulting from LBGs at these
redshifts have been explored by Adelberger \etal (2002), Croft \etal 2002,
and Kollmaier \etal 2002 (the latter two using simulations of
structure formation). 
In Croft \etal (1999), a simple analytic argument was made, using the
results of Kovner and Rees (1989) to show that to a very good approximation 
we can treat the radiation coming from these numerous sources as uniform.
At higher redshifts, as the opacity of the IGM increases, the number of 
sources within an attentuation volume decreases so that the fluctuations
can become large, (see e.g., Meiksin and White 2003b).
In our present simulations, we are concerned with $z=3$, and treat
the additional radiation intensity from these sources as 
coming from a uniform background
radiation field.

The mean intensity of this additional field is not observationally
very well constrained. 
Guided by the results of Sokasian, Abel and Hernquist (2003), we choose,
for our fiducial simulation
to add a uniform intensity equal to that of the QSO contribution.
We will also vary this quantity from a half to 
a factor of 64 times the QSO value in order to investigate its effects
on clustering of the \lya\ forest. We also directly compute
the mean intensity (at $912$ \AA\ )
contributed by the sources by first assigning the 
intensities seen by particles to a grid, and then volume
averaging. We find $J=2.4 \times10^{-22}$,
$1.7 \times10^{-22}$, and
$1.6 \times10^{-22}$
 for the d500, d250 and d125 simulations
respectively (in units of ${\rm erg s^{-1} cm^{-2} Sr^{-1} Hz^{-1}}$.)
 We assume a spectral index
for the uniform background identical to 
our assumed quasar spectral index of $\nu=-1$.
 This intensity can be compared to the value obtained using the 
proximity effect of 
$7.0^{+3.4}_{-4.4}\times 10^{-22} {\rm erg s^{-1} cm^{-2} Sr^{-1} Hz^{-1}}$ 
by Scott \etal (2000). The value of $J$ required to reproduce the
mean opacity of HI in the \lya forest
was estimated to be $2.3 \times 10^{-22}$ by Rauch \etal 1997 
(assuming an HM96 spectral shape; see also Weinberg \etal 1997).
We note that for the smaller of our box sizes, many of the photons
are not absorbed before they reach the maximum travel distance 
of half a box length. Because of the inverse square law, however, this only
results in a small difference in the mean flux between the simulations, as
mentioned above. We correct this difference by adding a small
extra uniform component to the intensity for the d125 and d250 simulations to
correct for the missing intensity, by comparing to the results of the
d500 simulation.

\subsection{The radiation field}
\begin{figure}[t]
\centering
\psfig{file=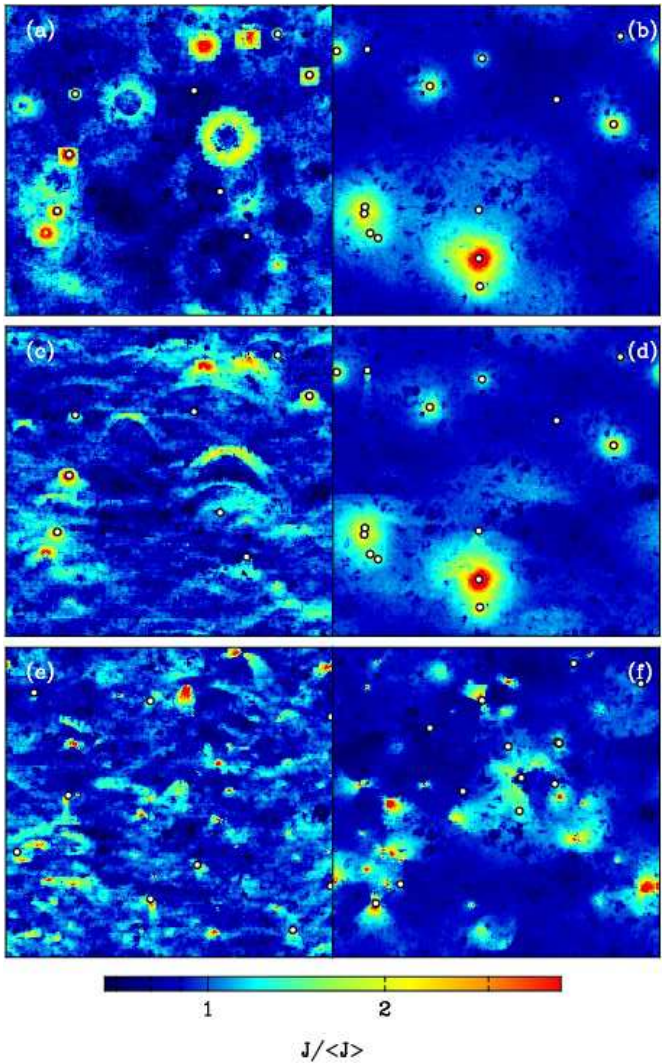,angle=-90.,width=8.75truecm}
\caption[]{ The radiation
intensity field ($J$, in units of the mean)
 in $25 \hmpc$ thick slices through one of the d250
simulation volumes (box side length $250 \hmpc$) at $z=3$. 
All panels show the same region of space, but with different assumptions/
approximations for the radiation emitted from quasars.
The left panels are for quasars with a lifetime
$t_{q}=10^{7}$ yrs and the right panels for 
$t_{q}=10^{8}$ yrs. Panels (a) and (b) are not output on the lightcone
and have no quasar beaming. Panels (c) and (d) include lightcone effects
(the observer's line of sight is the y-axis)  and 
isotropically emitted radiation. In panels (e) and (f), radiation
is emitted from quasars  in a cone of opening angle 90 degrees, and 
lightcone effects are included (see text). Quasars visible to an observer
are shown as points.  
  }
\label{slices}
\end{figure}

In Figure \ref{slices} we plot the
radiation field in slices through the simulation box, for one
of the d250 realizations. 
This gives an idea of both the scale and morphology of the 
individual features, as well as the large scale structure in the field.
The panels
on the left side are for $t_{q}=10^{7}$ yrs and those  on the right for
$t_{q}=10^{8}$ yrs. The top panels do not include light cone effects,
but the middle and bottom panels do, with the bottom panel being
computed for anistotropically emitted radiation. 
In all cases, we have added our fiducial uniform  background
radiation with a level equal to the QSO contribution, as mentioned above.

For the top panels, without lightcone effects, the light
echoes have the expected spherical shape. 
We show quasars visible (active) to an observer as points.
It can be seen that for the 
shorter-lived quasars, as expected, much of the radiation has been
emitted by quasars which are currently not active, whereas for $t_{q}=10^{8}$
yrs, many of the obvious high intensity radiation spots are centered on
visible sources. Because of the coherence over long timescales of the radiation
from the longer lived objects, the radiation field in the right panels
appears smoother on small scales. We shall see below when we quantify the
clustering in the Lya forest brought about by this inhomogeneous field
that it is stronger on large scales.

In the middle row of Figure \ref{slices},
the observer direction is towards the bottom of the plot, along the y-axis.
Because of this, as explained in \S 3.3 above, the light echoes are not
spheres. Quasars that are still on when the observation takes place 
have radiation around them which has propagated further in the
direction of the observer than away. Quasars which have switched off
recently enough that the inverse square law has not yet 
washed out their radiation are responsible for the bow shaped light echoes.

In the bottom row, the beamed quasars give the radiation field 
a different morphology. Roughly three times as many quasars are present
as in the other panels, but only those for which the observer's line of sight
is within the radiation cone opening angle are shown as points.
As in the other panels, most of the volume of space has an intensity
about $80\%$ of the cosmic mean, with the region of space with intensity
a factor of four or higher being very small indeed. The absorption seen in
 \lya\ forest effectively samples the volume of the Universe reasonably
fairly so that  these small excursions of high intensity are unlikely to
have much effect on \lya\ forest clustering.

\begin{figure}[t]
\centering
\psfig{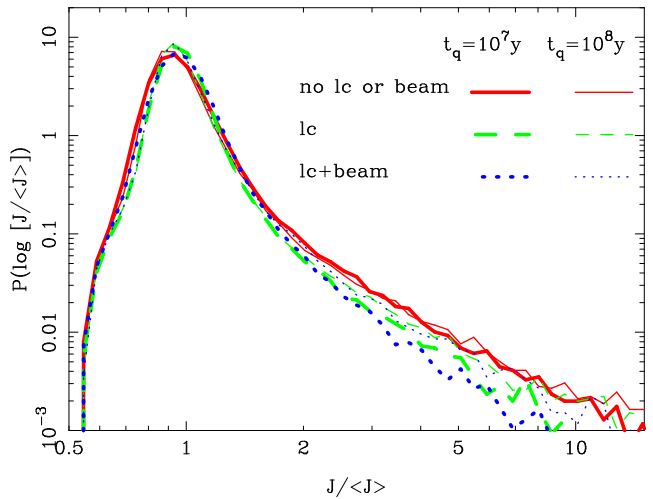}
\caption[radpdf]{ 
The PDF of radiation intensities in units of the mean
($J/\langle J\rangle$) in the d250 simulation
at $z=3$. We show results averaged spatially 
in cubes of side length $3 \hmpc$.
The y-axis gives the probability of finding a value of
$\log J/\langle J\rangle$  between
$\log J/\langle J\rangle$ and
 $\log J/\langle J\rangle+ d \log J/\langle J\rangle$.
The different line thicknesses show results for two differemt quasar 
lifetimes, and the different line styles results for either no beaming
or light cone (lc) effects, light cone effects only, or both (as in 
Figure \ref{slices})

\label{radpdf}
}
\end{figure}

In order to make comparisons between the panels more quantitative, we
have computed histograms of radiation intensities in cubical grid cells
of side length $3\hmpc$. The results are shown in Figure \ref{radpdf}.
In this plot we also indlude our fiducial
uniform background contribution as was done for the plots of slices
through the radiation field.
The $\%$ of pixels with $J$ twice the mean or more is $0.9-1.5\%$, and is
lowest for $t_{q}=10^{7}$ years with lightcone effects and
greatest for $t_{q}=10^{7}$ years without lightcone.  
The standard deviation of values  is greatest (1.1) 
for the beamed quasars with $t_{q}=10^{8}$ yrs and least (0.58) for 
beamed  $t_{q}=10^{7}$ years sources without lightcone effects.
On the small scale of these  cells, the difference
between the quasars with different lifetimes is slight. The 
$t_{q}=10^{7}$ years objects have slightly larger
fluctuations and a more skewed PDF, as might be 
expected from  the  complexity evident in the left panels of 
Figures \ref{slices}.
There is also even less change evident in the results when adding 
in lightcone effects and beaming.

\subsection{Lyman-alpha spectra}
Now that we have an inhomogeneous radiation field, we create \lya\
spectra that would be observed in quasar spectra whose sightlines pass
through it. We assume that the hydrogen is in photoionization
equilibrium with this radiation field (as mentioned previously, this 
is a good approximation if $t_{q} \simgt 10^{4}$ yrs). The neutral hydrogen
density is inversely proportional to 
the photoionization rate $\Gamma$:
\begin{equation}
\end{equation}
where 
\begin{equation}
\Gamma=\int^{\infty}_{\nu_{HI}}d\nu\frac{4\pi J(\nu)}{h\nu}\sigma_{HI}(\nu)
\end{equation}
The optical depth to absorption is given by 
\begin{equation}
\tau_{Ly\alpha}=\frac{\pi e^{2}}{m_{e}c} f \lambda H^{-1}(z)n,
\end{equation}
where $f=0.416$ is the \lya\ oscillator strength, and $\lambda=1216$ \AA\
 (Gunn and Peterson 1965)
 The neutral density $n$ is inversely proportional to the photoionization
rate for gas in photoionization equilibrium,  and hence $\tau_{Ly\alpha}$
is inversely proportional to the intensity
of the radiation field.
With high resolution simulations which resolve
the relevant length and mass scales (see e.g., Bryan \etal 1999 ),
the optical depth can be calculated directly for each point in the spectrum.
The optical depths are then convolved with the
line-of-sight peculiar velocity field and thermal velocities to 
produce a redshift space spectrum (see e.g., Hernquist \etal 1996).

Dark matter only simulations have also been used to make spectra
(e.g., Petitjean \etal 1995, Gnedin \& Hui 1998, Croft \etal 1998,
 Meiksin and White 2001), making use
of the fact that dark matter traces the gas density well on scales
larger than the Jean's scale.
In the present case, because we are are interested in large-scale 
clustering of the \lya\ forest, we have too low resolution to make 
realistic spectra directly, even using the dark matter as a proxy for
baryons. Instead, we again appeal to the high resolution hydro simulation
of \S 2.2.
Our approach involves finding the flux in relatively large pixels
which are associated with a given density of dark matter.

First we find the redshift space dark matter density and the
ionizing radiation intensity in the dark matter simulations in cubes
lying along sightlines through the box. The cubes are chosen to be relatively 
large (diameter $3/256$ of the box size, which is $3 \hmpc$ for the 
d250 runs). We compute redshift space quantities, shifting the 
particle positions along the sightline axis by adding their
peculiar velocities to their Hubble velocities.
As $J$ values are associated with each particle,
we compute the $J$ field in each cell from the particles which fall into the
cell in redshift space.

In order to make use of the high resolution hydro simulation, we then
randomly place within it cubical cells of the same size used
for the dark matter simulations and compute the
redshift space dark matter density in those cells. We next compute
1500 high resolution \lya\ spectra which pass through the centers of the cells
The spectra are computed from the neutral hydrogen density by integrating
through the SPH kernels  of the particles, and then convolving with the
line of sight velocity field, in the usual manner (see e.g., Hernquist
\etal 1996, or Croft \etal 2002 for the same simulation).
The intensity of the ionizing background can be adjusted after the simulation
was run, and we do so to reproduce a mean \lya\ forest flux of
  $\langle F\rangle=0.64$ (from Press, Rybicki and Schneider 1993).
We then average the unabsorbed flux
 $\langle F\rangle=\langle e^{-\tau}>$ inside each pixel that 
lies onspectra passing through the center of each cell, so that we have 
the F value which would be
observed for a large pixel of size 32, 65, and 130 $\kms$ 
for the d125, d250 and d500 simulations respectively
(the cell size in redshift space).
We make spectra several times over, each time varying
the intensity of the ionizing  background, from 100 to 0.01 times our 
fiducial intensity, in 20 log spaced
bins.

Given this information, we now have, for a given overdensity of 
dark matter and radiation intensity at 912 \AA\ 
 a distribution of possible values for
the \lya\ forest flux in a large pixel. We use this to map our large 
low resolution dark matter skewers into spectra. We try two different 
methods for doing this. The first method (which we shall use for the rest of
the paper) is to pick a random F value from the
set of pixels with the correct dark matter density and J value. The second
is to assign to the pixel the mean value of $\langle F\rangle$ for all pixels with the
correct correct dark matter density and J value. For all the statistics
we compute in the rest of the paper there is negligible difference
between the two methods.

\begin{figure*}
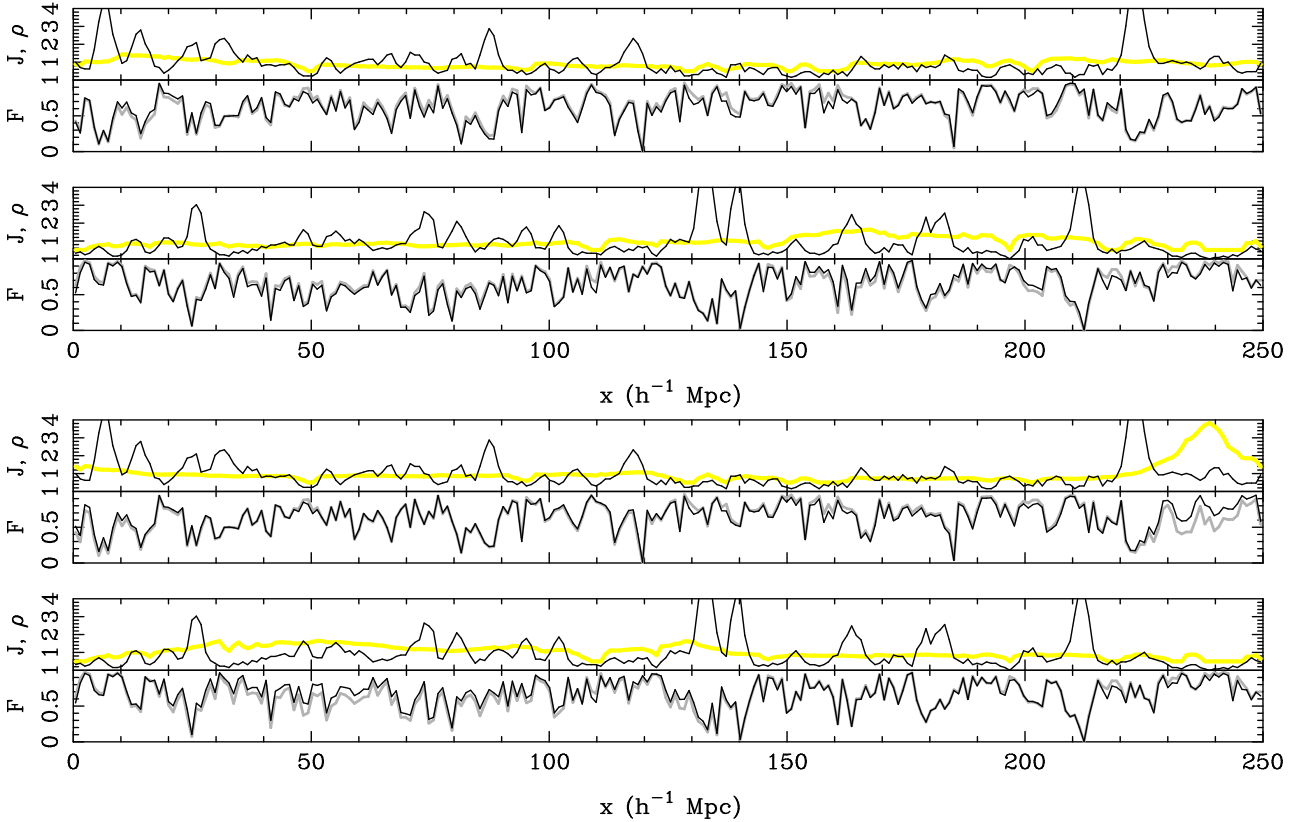

\begin{centering}
\psfig{file=spec1e7.ps,angle=-90.,width=17.0truecm}
\psfig{file=spec1e8.ps,angle=-90.,width=17.0truecm}
\end{centering}
\caption[]{ 
Two randomly chosen line of sight density (dark line) and radiation
(light line) fields
 (in units of the mean) for 
the no lightcone $t_{q}=10^{7}$ yr quasars (top 2 panels) and
for the no lightcone $t_{q}=10^{8}$ yr quasars (bottom 2 panels).
Under the radiation and density we show spectra made from these sightlines
with a uniform radiation field (light line) and non uniform (dark line).
}
\label{los}
\end{figure*}

In Figure \ref{los} we show the \lya\ spectra that result from two randomly
chosen sightlines through one of the d250 simulations. We also show the dark
matter density and radiation intensity along the same sightlines, both in
units of the cosmic mean. Each sightline is shown twice, for two 
values of the quasar lifetime ($t_{q}=10^{7}$ yrs and $t_{q}=10^{8}$ yrs)
and results are without lightcone effects.
The density field (smoothed with a $3 \hmpc$ filter) obviously
fluctuates more than the J field, showing that the density is the source
of most structure. There is only one moderately
 large excursion in $J$ above the mean
for the first sightline (for $t_{q}=10^{8}$ yrs) where a foreground quasar was
present. The difference between the two values of $t_{q}$ is not
easy to see in the radiation field, but it seems plausible that 
there are larger scale $J$
fluctuations for longer $t_{q}$. The \lya\ forest spectra are shown next 
to spectra that result from a uniform $J$. The difference is very small,
with a maximum difference in $F$ of 0.17 between the two. The effect
of gentle long wavelength fluctuations in $J$ can be seen, particularly
in the bottom-most panel.

\section{The Lyman-alpha power spectrum}

Having created an inhomogenous radiation field from a population
of quasar sources and \lya\ forest spectra running through it, we now
turn to \lya\ forest observables. In this section we measure the large-scale
clustering of our \lya\ forest spectra using the power spectrum, and 
investigate its dependence on the assumptions we have made 
regarding quasar lifetimes, radiation contributed by other sources and
inclusion of lightcone effects and beaming. The power
spectrum of the density distribution in the Universe
is a useful probe of cosmology, and the relatively simple
physics relating \lya\ forest flux and density 
has made it possible to use \lya\ forest data for this 
purpose (see e.g. Croft \etal 1998, Hui 1999, McDonald \etal 2000). 
Observational constraints on P(k) have been made by 
McDonald \etal (2000), Zaldarriaga, Hui \& Tegmark (2001),
 Gnedin \& Hamilton (2002) and Croft \etal (2002), amongst others. 
It is of particular interest to see if radiation fluctuations
can disrupt the relation between density and \lya\ forest flux
enough to affect cosmological measurements.

We compute the one-dimensional power spectrum of the \lya\ forest
spectra by first calculating a mean flux contrast field:
\begin{equation}
\delta_{F}(\lambda) \equiv F(\lambda)/\langle F \rangle-1
\end{equation}
The power spectrum is then given from the variance of Fourier modes
with a given $k$, so that
\begin{equation}
\delta(k)=\frac{1}{2\pi}\int\delta(x)e^{-ikx}dx,
\end{equation}
and the dimensionless (1-dimensional) power spectrum,
\begin{equation}
\Delta_{F}^{2}(k)=\frac{k}{\pi}\langle \delta(k)^{2} \rangle.
\end{equation}
In Figure \ref{numbertest},
 we show some simulation results, which we shall describe below.
We also show observational results for $\Delta^{2}_{F}(k)$ from Croft \etal 
(2002). These were computed from pixels in spectra which were within
$\delta_{z}=^{+0.2}_{-0.1}$ of $z=3$. The length scales 
in $\kms$ have been converted to
comoving $(h^{-1} Mpc)^{-1}$ assuming our $\Lambda$
cosmology. The half-wavelength $\pi/k$ of the 
largest scale plotted is $14 \hmpc$. This measurement was made from a
subsample of the Croft \etal dataset which contained 53 quasar spectra.
Datasets two or three orders magnitude larger are forthcoming from
the Sloan Digital Sky Survey (see e.g., Hui \etal 2003, Seljak \etal 2003),
and it will be possible to measure $\Delta^{2}_{F}(k)$ on much larger 
scales with them (modulo other uncertanties such 
as continuum fitting).

\subsection{Convergence tests}
\begin{figure}[t]
\centering
\psfig{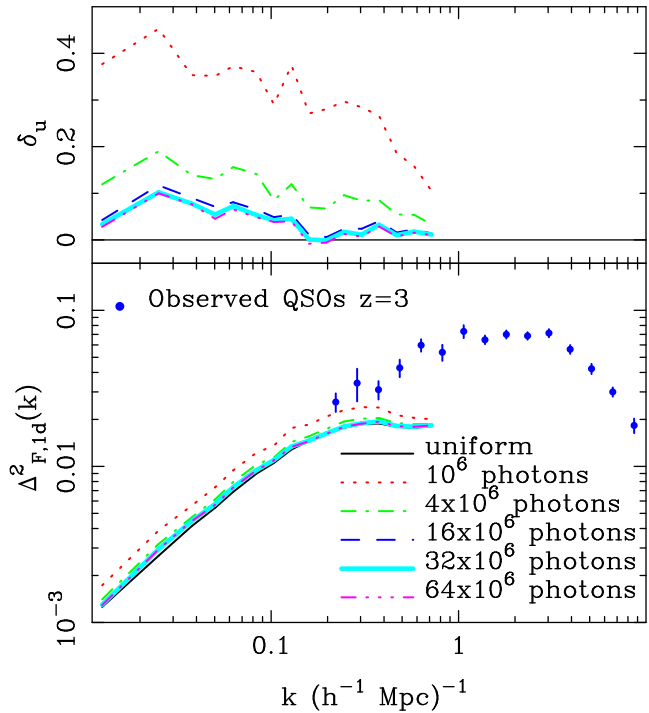}
\caption[numbertest]{ 
\label{numbertest}
Convergence test for the number of photon packets to be used in raytracing. We
show the power spectrum of the \lya\ forest flux $\df$ in the d500 simulation
(no light cone effects or beaming), for different numbers of photon packets.
The top panel shows the fractional difference
in $\df$  between the raytraced simulation and
one using a uniform radiation field (Equation \ref{deltau}). 
}
\end{figure}

With our Monte-Carlo method for raytracing, shot noise in the number
of photon packets could result in unphysical fluctuations in the
radiation field if not enough are used. 
In Figure \ref{numbertest}, we compute $\df$
using 1000 \lya\ spectra taken from a single realization of a d500 simulation. 
In the bottom panel, $\df$ itself is plotted, whereas in the top panel we
show the ratio of $\df$ for spectra computed with a uniform background
which we call 
\begin{equation}
\delta_{u}=^{\rm uniform}\df/\df.
\label{deltau}
\end{equation}
We compute these quantities after using a variable number of
photon packets
 to generated the radiation field, from $10^{6}$ to $64 \times 10^{6}$.
The results appear to have converged after $16 \times 10^{6}$ photon
packets. For
the rest of our calculations, we shall use $32 \times 10^{6}$.

\begin{figure}[t]
\centering
\psfig{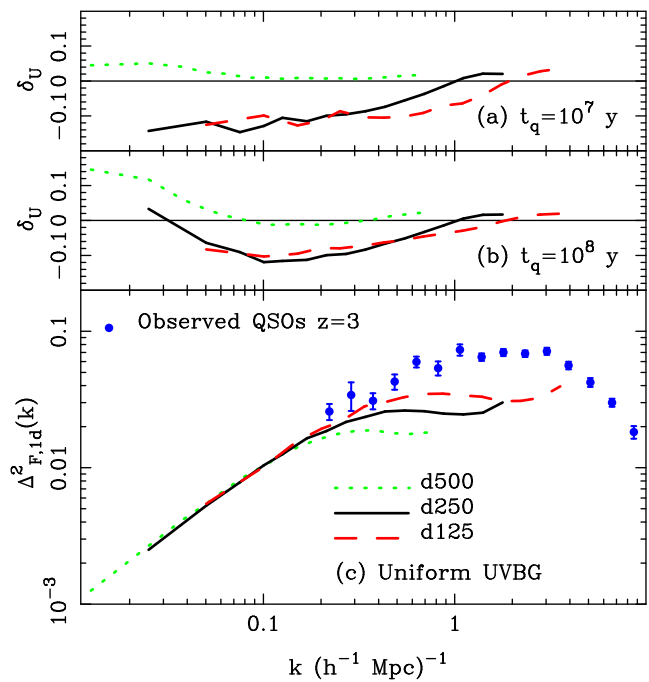}
\caption[conv]{ 
\label{conv}
Convergence test for the box size, force resolution and particle mass. We
show the power spectrum of the \lya\ forest flux $\df$ averaged over
all realizations for the three sets of simulations with different
box sizes and particle masses. In this case output was on the
lightcone, but the radiation output of quasars was set to be
isotropic.
The top panel shows the fractional difference (Eqn. \ref{deltau})
in $\df$  between the raytraced simulation and
that using a uniform radiation field. 
}
\end{figure}

As we have run simulations with three different box sizes, we are also 
able to check convergence of $\df$ with spatial resolution and particle mass.
For the rest of the figures in this paper, we use 1000 spectra from
each simulation realization and average over 5 realizations.
 In Figure \ref{conv} we show $\df$
measured from the d125, d250 and d500 simulations (all outputs including
lightcone effects, but no beaming). The bottom panel
shows results for a uniform radiation field,
 and the top two the ratio $\delta_{u}$
for the two different lifetimes. In the bottom panel we can see that
the absolute value of $\df$ has converged well over the range of scales 
that are resolved by all 3 simulations ($\sim 0.05-0.15 \invhmpc$).
The d125 simulations are of high enough resolution that they can be compared
with one or two of the largest scale observational points, and are
reasonably consistent.

In the top two panels of Figure \ref{conv}, we can see that the 
power spectra relative to that with a uniform $J$ only converge
when we reach the d250 resolution. The d500  $\delta_{u}$ is actually
slightly greater than 1, whereas the converged result shows that the
$J$ fluctuations suppress the \lya\ power spectrum (as seen by 
Meiksin and White 2003b at higher redshifts $z>4$). The power spectrum
suppression is about $10-12 \%$ on scales $k \sim 0.1 \invhmpc$. Both
values of $t_{q}$ give similar results on these scales, whereas on 
larger scales $k \sim 0.05 \invhmpc$, $\delta_{u}$ for the long lived
quasars is becoming positive again. This corresponds to
physical scales of half-wavelength $\sim 100 \hmpc$, or
 close to the distance
light can travel in $10^{8} yrs$. For scales $k \simgt 1 \invhmpc$, 
$\delta_{u}$ becomes close to zero, at least for the large value of $t_{q}$,
indicating that fluctuations due to the inhomogeneous  radiation
field are unlikely to affect $\df$ on presently observed length scales at
more that the $1-2 \%$ level. For the results with $t_{q}=10^{7}$,
the results have not converged quite as well on the smallest scales,
indicating that there is room for a small amount of excess power. However, 
this regime of lengthscales has been probed (at $z=4$)
by Gnedin and Hamilton (2002), and  by Meiksin and White (2003b), who find
minimal effects.

\subsection{Lightcone, beaming and uniform background
effects on the \lya\ power spectrum}

\begin{figure*}
\centering
\hspace{0.1cm}\psfig{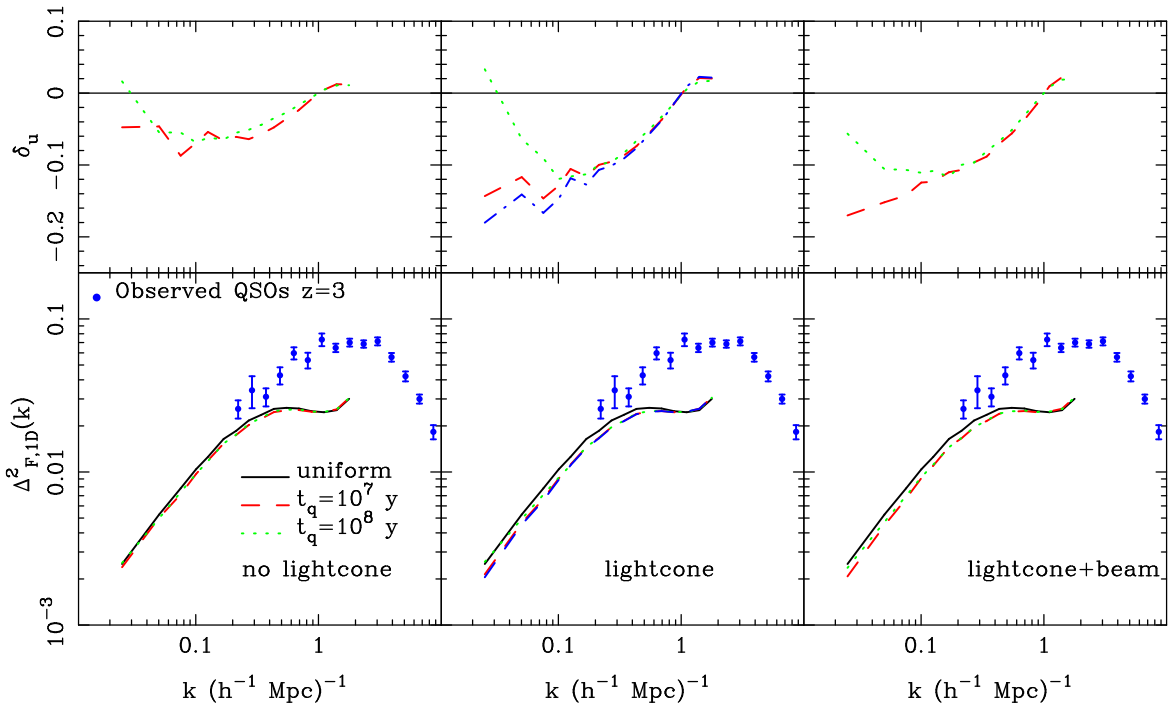}
\caption[]{ 
The effect of output on the lightcone and anisotropic QSO radiation
on the power spectrum of the \lya\ forest flux $\df$. We show results
for a uniform radiation field, and raytraced field with
two quasar lifetimes. In the middle panels, the dot-dashed line is the 
result for the uniform attentuation approximation (see text).
The top panels show the fractional difference
in $\df$  between the raytraced simulations and
those using a uniform radiation field. 
  }
\label{fpkratio}
\end{figure*}

Given that the d250 simulations have converged for most of the
range of length scales we are interested in, we now examine the
effect of including output on the lightcone and anisotropically
emitted radiation in these runs. In Figure \ref{fpkratio} we show 
$\df$ for outputs with and without these effects, for both values of $t_{q}$.
We find that including lightcone effects increases the amplitude of
the radiation fluctuations, making suppression of the \lya\ power 
spectrum stronger by roughly a factor of 2. The presence of beamed
radiation makes little difference, however, which is somewhat suprising.
It seems as though the smoothing effect on the radiation field
from having more sources per unit volume has been cancelled  
out by the fluctuations due to the limited beaming angle around each source.
Even though the effect of beaming on the one dimensional \lya\ forest
statistic is minimal, the radiation field itself does look 
dramatically different, as one can see by comparing panels (e) and (f)
of Figure \ref{slices}.

The results of an additional test are shown in Figure \ref{fpkratio}.
As mentioned in \S 3.2 we have also computed the radiation field assuming 
the uniform attentuation approximation, where there is no shadowing of
light and photons travel isotropically from each source. We have
used an attenuation length of $111 \hmpc$ appropriate for a
uniform medium. For the lightcone case with $t_{q}=10^{7}$ y, we
plot the results in Figure \ref{fpkratio},  where they
can be compared with the full
treatment. We find that the differences are quite small, 
with $\du$ being essentially identical for $k >0.05 \invhmpc$
and only different by $20\%$ on the largest scales. From this we 
can conclude, that the effect of shadowing in the largely optically
thin Universe at $z=3$ is minimal. For this statistic at least, the
inclusion of lightcone effects is more important.

\begin{figure}[t]
\centering
\psfig{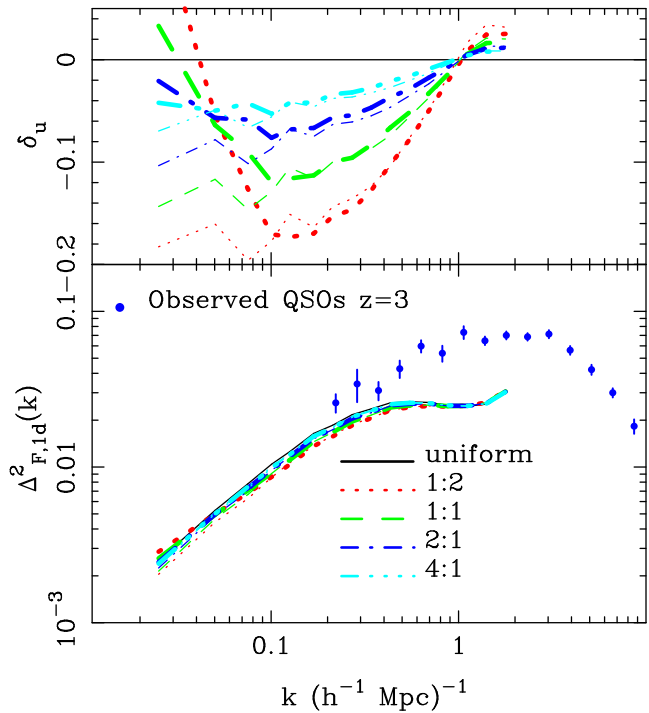}
\caption[bglevel]{ 
\label{bglevel}
The effect of adding a uniform component to the 
raytraced radiation field
on the power spectrum of the \lya\ forest flux $\df$. 
The ratio of uniform component to 
component coming from QSO sources is shown in the legend (e.g., 2:1 means
that the contribution to the spatially averaged intensity from a uniform
field is twice that from the sources). 
We show results for the d250 simulations, with output on
the lightcone, and for $t_{q}=10^{7}$ yrs (thin lines), and
 $t_{q}=10^{8}$ yrs (thick lines)
The top panel shows the fractional difference
in $\df$  between the raytraced simulations and
those using a totally uniform radiation field. 

}
\end{figure}

As explained previously, there will be a much more uniform component to the 
radiation field, originating from recombination radiation and stellar
sources in galaxies. We expect that the mean intensity contributed in this way
will be approximately equal to that from quasars (Sokasian \etal 2003). 
In Figure \ref{bglevel}, we explore the effect on $\df$ of adding an
an extra uniform contribution. We vary this uniform background to be from
0.5 to 8 times that of amount from quasars, and find that the suppression
of $\df$ relative to a uniform $J$ field goes down as the uniform background
level is raised, as expected. Doubling the uniform field compared to
our fiducial level changes the maximum supression from $12 \%$ to $8\%$
for $t_{q}=10^{8}$. With a uniform field larger than our fiducial one,
there is no boost in $\df$ on the largest scales either.
We have also tried reducing our fiducial uniform background by a factor of
2, and results are also shown in Figure \ref{bglevel}. In this case, the
suppression measured by $\du$ is stronger (about $18\%$ maximum suppression).

It is interesting to examine the potential effects of this suppression
of $\df$ on work which seeks to constrain the matter power spectrum
from \lya\ forest measurements. Data from the Sloan Digital
Sky Survey will probe larger scales than those in the analyses
by Croft \etal (2002), for example. A uniform suppression of $\df$ by
$10 \%$ at redshifts close to these  would translate to a reduction in the
inferred amplitude of mass fluctuations ($\propto \sigma_{8}$) of $\sim 7 \%$
(Croft \etal 2002).  The fact that $\du$ changes with scale
will slightly alter the shape of the inferred mass power spectrum.
We will return to this in our discussion in \S 6.

\section{ Foreground proximity effect}

We have seen that the effect of radiation fluctuations on the overall
clustering of the \lya\ forest at $z=3$ is likely to be small. Close to QSO
sources, however, the locally produced radiation can overwhelm
the background and produce much stronger effects. The effect of a QSO's 
radiation on its own \lya\ forest, known as the proximity
effect, has been extensively studied, first by Bajtlik, Duncan and
Ostriker (1988). Because the ionizing radiation studied in such a case is
that along the line of sight to the quasar, as long as the quasar source
turned on within the equilibration time ($\sim 10^{4}$) years of being 
observed, the \lya\ forest will respond and the effect can
be seen. The relative suppression of
absorption (quantified by the number of \lya\ lines) caused by the ionizing
radiation close to the quasar has been used to constrain the overall
intensity of the background field by Cooke, Espey \& Carswell (1997),
Scott \etal (2000), among others.

The effect of radiation from a QSO on the \lya\ forest of other quasars
is known as the foreground or transverse proximity effect. Because
the light travel time to a sightline 10 $\hmpc$ away in the 
transverse direction is $\sim 10^{7}$ y, the presence or
absence of an effect can be used to constrain the lifetime of the
QSO. The use of the foreground proximity effect to do this has been 
explored recently by Schirber, Miralda-Escud\'e, \& McDonald
 (2003), using data from
the Sloan Digital Sky Survey Early data release. These authors found
no evidence of an effect, something we shall return to later.
The foreground effect has also been examined theoretically by e.g., Kovner
 \& Rees (1989), and possible detections have been reported 
by Dobrzycki \& Bechtold (1991), Jakobsen \etal (2003),
 and limits by Liske \& Williger (2001).

In this paper, we will not examine the foreground proximity effect
around individual sources but instead the averaged \lya\ forest
flux around all quasars,
$\langle F\rangle(r)$. The mean \lya\ forest flux as a function of 
distance from Lyman Break Galaxies was measured by Adelberger \etal 
(2003), and from simulations by Croft \etal (2002), and Kollmaier \etal (2002),
and Bruscoli \etal (2003).
We will carry out the same measurement for QSOs, making predictions from
our simulations and an observational measurement from SDSS data. We will
not be measuring the effect of each quasar on its own \lya\ forest,
as this requires accurate modelling of the continuum close to
the quasar, including its \lya\ line emission, and is left for future work.

\subsection{SDSS First Data Release}
Our observational determination of $\langle F\rangle(r)$ is made using the first data
release of the Sloan Digital Sky Survey 
(Abazajian \etal 2003). We have taken all spectra
for which some part of the \lya\ forest region is available above
$z=2.2$ which includes 1920 spectra.
The mean redshift of all \lya\ forest pixels in this region is $z=2.81$,
and the mean S/N is 5.7. We do not fit a continuum to these spectra but
remove the continuum dependence by instead  first 
smoothing them with a Gaussian filter of width $\sigma=50$ \AA\. We 
do not try to determine the absolute 
mean flux level of the spectra, but  assign a value based on
the work of Press, Rybicki and Schneider (1993, PRS),
 which is largely consistent
with the mean flux measured from SDSS spectra by Bernardi \etal (2002).
To do this, the flux value of a pixel at redshift $z$ is 
given by
\begin{equation} 
F_{z}=\frac{S_{z}}{C_{z}}\times PRS_{z},
\label{cont}
\end{equation}
where $S_{z}$ is the spectrum, $C_{z}$ is the value of the
smoothed spectrum and $PRS_{z}$ is the value of the mean flux
from the PRS data at redshift $z$. We note that because our simulated spectra
were normalized to have the PRS mean flux we are therefore
consistent. Continuum fitting is a difficult procedure and uncertainties 
in the observationally determined values are still large. For example, it
is very possible that the PRS determination of the mean flux is too low (see
e.g., Seljak, McDonald \& Makarov 2003)
. Although this is an important question which
must be resolved, in the case of the present paper, as long as the simulations
and observations are normalized using the same values, these
uncertainties will not affect our comparisons.

As mentioned above, the region close to the \lya\ emission line must be
treated with special care if it is to be used. In the current work, we
are conservative 
and instead only use \lya\ pixels with
 rest frame wavelength between $1060$ \AA\
 and
$1160$ \AA\. Also, as we are interested in the mean flux of pixels
averaged around QSOs, only pixels which fall close to QSOs will actually
be used. We convert redshifts and angular coordinates of all pixels into
3 dimensional cartesian comoving coordinates, assuming a flat cosmology with 
$\Omega_{M}=0.3$ and $\Omega_{\Lambda}=0.7$. After doing this, only
pixels within 40 $\hmpc$ of a QSO are kept. This cuts down the number
of contributing QSOs to 325. The mean $z$ of these pixels is $z=2.6$,
and the mean absolute QSO magnitude in the SDSS $G$ band is $-27.2$ 
(our simulated quasars have a similar mean $G$ magnitude of $-27.0$, again
assuming a spectral index of $\nu=-1$ to transform
from $B$ to $G$ magnitude).
 Because we are interested in comparing to our simulation results at 
$z=3$, we have changed the mean value of the absorption in Equation 
\ref{cont} to $PRS_{z+0.4}$. By doing this, we are assuming that the
effect of changing mean flux between $z=2.6$ and $z=3.0$ 
is more important to $\langle F\rangle(r)$ than the change in
the density field around QSOs. In order to test this, we have tried
setting the lower bound on pixel redshifts to $z=2.5$, which gives
a mean pixel redshift of $z=3.0$, and calculating 
$\langle F\rangle(r)$ without using any offset in Equation \ref{cont}. We find
results that are much noiser, but consistent within the errors. With future 
releases of SDSS data it will be possible to
measure $\langle F\rangle(r)$  of a larger 
sample of QSOs at many redshifts.

\begin{figure}[t]
\centering
\psfig{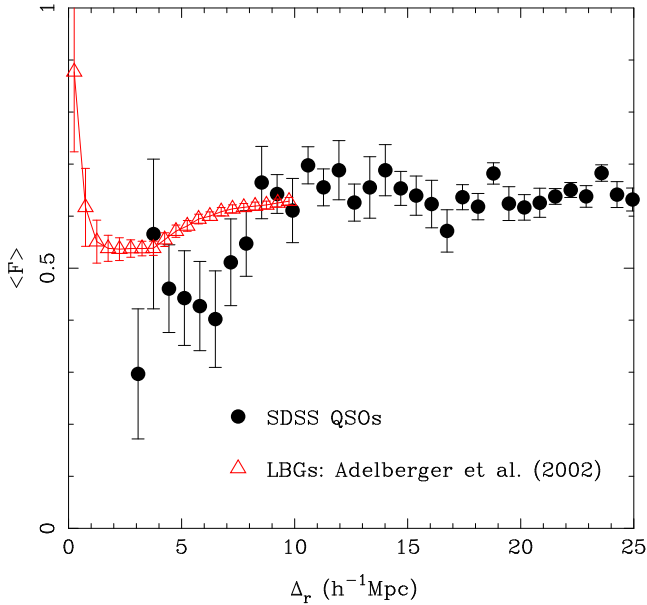}
\caption[kurt]{ 
\label{kurt}
The mean \lya\ forest flux $\langle F\rangle$ at $z=3$ averaged around quasars
in the SDSS First Data Release as
a function of quasar-pixel distance (round points). Also 
shown are results  at $z=3$ for Lyman-break galaxies (Adelberger \etal 2002).
}
\end{figure}

In Figure \ref{kurt}, we show $\langle F\rangle(r)$  for the SDSS QSO sample,
 compared
to the results for Lyman Break Galaxies of Adelberger \etal (2002). The
Adelberger \etal points show increased absorption with respect to the mean
over the range from $\sim 2 -10 \hmpc$ (comoving)
from galaxies. On smaller scales,
there is evidence (at the $2-3 \sigma$) level for decreased absorption, or
a proximity effect. What the cause of this lack of absorption
close to galaxies might  be was considered extensively by
Adelberger \etal (see also Croft \etal 2002 and Kollmeier \etal 2002). It 
was concluded that a radiation proximity effect could not be the answer
because the ratio of the ionizing background radiation to the galaxies
own is too high, even at distances of $\sim 1\hmpc$. 
Galactic winds blowing holes in the IGM seem to be a possibility,
although hydrodynamic simulations of winds in a cosmological 
context do not exhibit the required strong modulation
of the \lya\ forest flux (see Theuns \etal 2002, Viel \etal 2003). 
Measuring the redshift of the galaxy to enough precision
that any signal at these small scales is not washed out by position errors
is extremely difficult, so that Adelberger \etal caution that this
proximity effect result is tentative at best.

The space density
of the LBGs in present samples is
$1.5 \times 10^{-2} h^{3} {\rm Mpc^{-3}}$ (for a magnitude
limit of 25.5, see e.g., Adelberger \etal 2002).
This is a factor $\sim 1000$ times
 higher than that of the SDSS QSOs we have used to compute
the solid points in Figure \ref{kurt}. 
With $\langle F\rangle(r)$ for the QSOs, there
is also substantially more absorption than the mean, again up to distances
of around $\sim 10 \hmpc$. This excess absorption is to 
be expected if QSOs sample dense environments, and if the effect of
this excess absorption is not outweighed by the effect of ionizing 
radiation from the QSO. We can see that there appears to be no direct evidence
of ionizing radiation, no foreground proximity effect. In the next subsection,
we shall explore what we expect from our models and make a comparison. For
now, it is obvious that the excess absorption is substantially
greater than around LBGs at $z=3$, so that the QSOs appear to be located
in much denser environments and possibly more massive dark matter halos.
There is no information on scales below $3 \hmpc$ due to the 
sparsness of the QSO sample. Within $10 \hmpc$, 22 QSOs contribute
at least 1 pixel to the measurement of  $\langle F\rangle(r)$, but 
for the 3 points
with $r <= 5 \hmpc$, only 5 contribute. 
We note that these numbers are also equal to the numbers of pairs of QSOs 
with these separations.
This is because our restricted wavelength range means that only the 
lower redshift
QSO in each pair is used as a center when we average pixels
at a given distance from it.
The mean SDSS G magnitude of the
QSOs in both these cases is $-27.2$.  The error bars in Figure \ref{kurt}
have been computed using a jackknife estimator (Bradley 1982) with a number of
subsamples equal to the number of quasars in each bin.

\begin{figure}[t]
\centering
\psfig{file=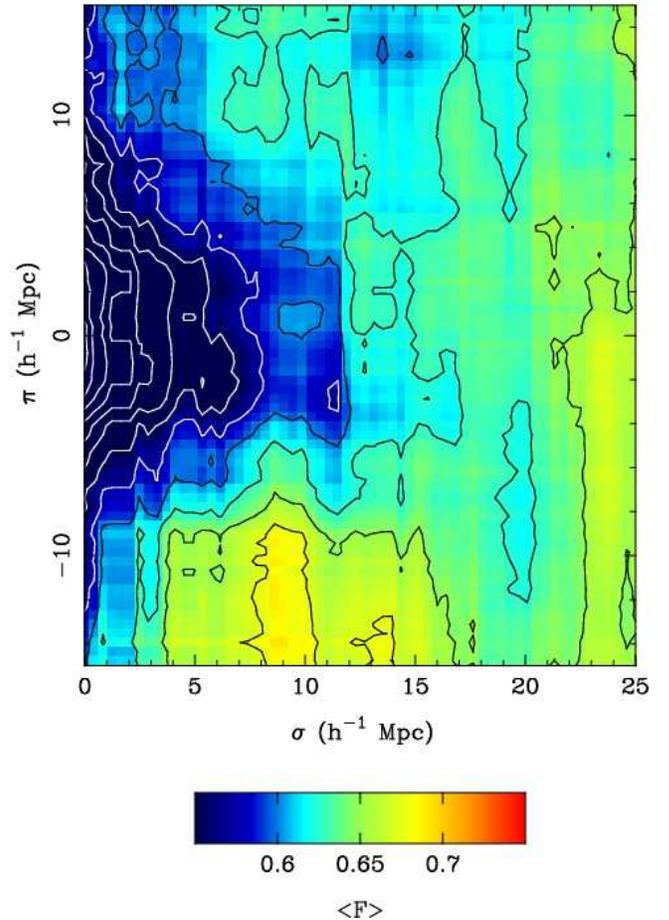,angle=-90.,width=8.5truecm}
\caption[]{ 
\label{sigpisdss}
The mean \lya\ forest flux $\langle F\rangle$ at $z=3$ averaged around quasars
in the SDSS first data release as
a function of quasar-pixel distance across the
line of sight ($\sigma$) and along the line of sight ($\pi$).
In order to reduce the effects of noise, the data was smoothed with
 a Gaussian filter with standard deviation $5 \hmpc$.
}
\end{figure}

The data shown in Figure \ref{kurt} is a radial average of the mean
flux around QSOs.
However, there are many reasons to suppose that the absorption could
show a different dependence on QSO-pixel distance transverse to (often 
referred to as $\sigma$)
 and along
the line of sight ($\pi$). We will investigate these in more detail when we
consider models below. For now, we will note that any ionizing radiation
propagating in the direction away from the observer (+ve $\pi$)
will have less time to reach a pixel at a given $|r|$ than if it were
moving in the opposite direction. This as we have already noted 
is the cause of the
lightcone effects seen in \S 3.3, and will lead to the mean 
flux $\langle F\rangle$ being possibly asymmetric about the $\pi=0$ axis. 
An additional possibility is that beamed radiation will cause a proximity
effect only at small angles to the $\pi$ axis. In order to investigate
these possibilities we have plotted $\langle F\rangle (\sigma, \pi)$ 
for our SDSS sample in Figure
\ref{sigpisdss}. In order to make the contours more visible, we
have smoothed the plot with a Gaussian filter with
standard deviation $_{e}\sigma 5 \hmpc$.
The excess absorption compared to the mean is clearly visible
close to the position of the QSO. There is also no obvious sign of asymmetry
about the $\pi=0$ axis. In order to 
quantify this, we have computed jackknife errors for
 $\langle F\rangle (\sigma, \pi)$ (in $5\times5 \hmpc$) bins. 
We have then computed the
sum 
\begin{equation}
\chi^{2}=\sum_{i,j} \langle F\rangle(\sigma,\pi)-\langle 
F\rangle(\sigma,-\pi)/_{e}\sigma^{2}
\end{equation}
where $_{e}\sigma^{2}$ is $_{e}\sigma_{i,j}^{2}+_{e}\sigma_{i,-j}^{2}$
is the error. We find
$\chi^{2}=60.6$ for 64 bins, so that within the statistical
uncertainty there is no evidence for any asymmetry. 
Figure \ref{sigpisdss} does not show very marked
elongation of the contours along the $\pi$ axis either, which could be
a sign that the QSOs are not found in large virialized systems.

\subsection{Simulation results}

\begin{figure}[t]
\centering
\psfig{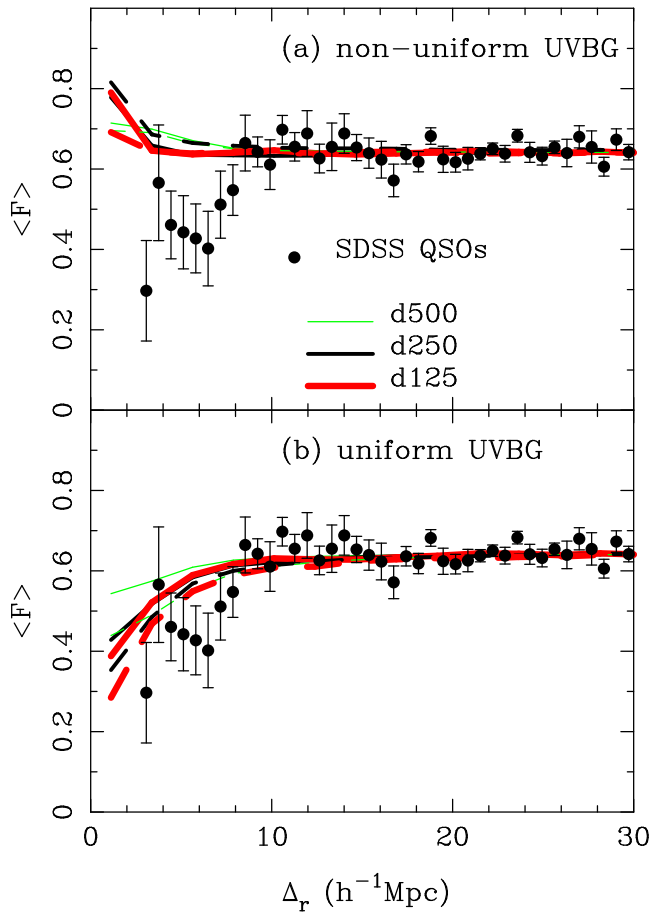}
\caption[favres]{ 
Effect of resolution and box size on the \lya\ forest flux
averaged around simulated quasars as a function of quasar pixel
distance. We show results for all three simulation sets, in each case
averaging over all realizations. The solid lines are for quasar lifetime
$t_{q}=10^{7}$ y and the dashed lines for $t_{q}=10^{8}$ y (output
is on the lightcone but there is no beaming of radiation).
The solid points are the observational data computed from the Sloan Digital 
Sky Survey First Data Release.
\label{favres}
}
\end{figure}

We have seen that the SDSS QSOs show no signs of decreasing the
absorption in other lines of sight passing close by. The situation
is rather different in our simulations, as we see in Figure \ref{favres}.
Here we have computed $\langle F\rangle(r)$ for all three sets of simulations,
the d500, d250 and d125, and for the two different
quasar lifetimes. In the top panel, we show results including
an inhomogeneous radiation field (with lightcone effects, but no beaming).
It is clear that there is a slight proximity effect close to the QSOs, and
that there is not much difference between the different simulations.
As with the convergence test of $\df$ (Figure \ref{conv}), the d250 and d125
runs appear to be somewhat closer together than the d500 results. The
simulation curves differ radically from the observational 
results on scales  $ r < 8 \hmpc$. This is rather surprising, as based
the simulations, we would expect to see some sign of the radiation
field in the real data, but there appears to be none.

Without QSOs acting as sources, however, we do find a closer match.
In the bottom panel of Figure \ref{favres}, we show the $\langle F\rangle(r)$
curves which result in our simulations if we assume a homogeneous radiation
field. The increased absorption resulting from QSOs being found
in dense environments is then seen, and it is clear that the almost
flat nature of $\langle F\rangle(r)$ in the top panel is the 
result of this increased
absorption being largely cancelled out by the QSO radiation flux. In the
uniform radiation field case, the QSOs with $t_{q}=10^{8}$ have 
marginally more absorption on scales $r <12 \hmpc$ from the QSO 
(this quantity has converged well with resolution for the d125 and d250 
simulations). If we compare with observations, though, even these rare
quasars do not have enough absorption to match well.

\begin{figure}[t]
\centering
\psfig{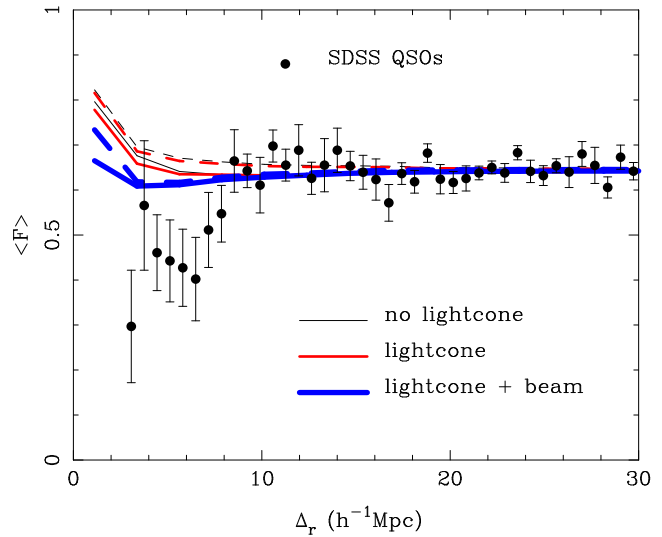}
\caption[fav]{ 
\label{fav} 
Effect of anisotropic
radiation and output on the light cone on the \lya\ forest flux
averaged around simulated quasars as a function of quasar pixel
distance. We show results for the d250 simulation set only,
averaging over all realization. The solid lines are for quasar lifetime
$t_{q}=10^{7}$ y and the dashed lines for $t_{q}=10^{8}$ y.
The solid points are the observational data computed from the Sloan Digital 
Sky Survey First Data Release.
}
\end{figure}

With the simulations, we can test the effect of turning off output on 
the lightcone. This is shown in Figure \ref{fav}. We might not
expect this to have much effect on this statistic because the
distances of interest correspond to relatively short light travel times.
From the plot we can see that the effect of ignoring lightcone effects
is neglgible for $t_{q}=10^{8}$ yrs, and not important
even for $t_{q}=10^{7}$ yrs. Of course if we were to 
condider $t_{q}$ of the same order or less as the light travel
time across the proximity effect region, then these
effects would become substantial. Schirber \etal (2003) have 
studied the proximity effect as a probe, across sightlines and conclude
that it can yield  interesting constraints on $t_{q}$. They find
that a $t_{q}$ of the order of $10^{6}$ years could be part of
the explanation for the lack of proximity effect seen by them in a sample
of SDSS data. We will return to this in \S 6.

If the radiation emission is highly anisotropic, then of course we would
not expect to see regions off axis being illuminated by the QSO. 
The radially averaged statistic $\langle F\rangle(r)$ should therefore show less of a
proximity effect, as we are averaging with some pixels at angles
which see only the background radiation. Figure \ref{fav} shows that this
is indeed the case, although the beaming angle we have chosen 
(opening angle of cone=$90\deg$) is too large to have much of an effect, and
$\langle F\rangle(r)$ in the simulations is still too high on small scales
compared to the observations. Our opening angle was chosen based
on what is considered theoretically  likely (Urry \& Padovani 1995).
 The fact that
an opening angle much smaller (by a factor of 3 or more)
would be necessary in order to even
approach what is seen in the observations makes it likely that anisotropically
emitted raditiaion is not the sole cause of the discrepancy
between our simulations and observations.

\begin{figure}[t]
\centering
\psfig{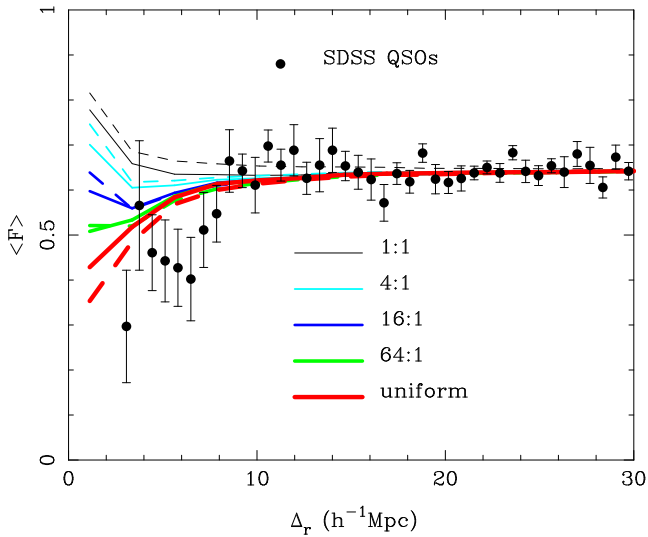}
\caption[favuniform]{ 
\label{favuniform}
The effect of adding a uniform component to the 
raytraced radiation field
on  the \lya\ forest flux
averaged around simulated quasars as a function of quasar-pixel
distance.
The ratio of uniform component to 
component coming from QSO sources is shown in the legend (e.g., 16:1 means
that the contribution to the spatially averaged intensity from a uniform
field is 16 times that from the sources). 
 We show results for the d250 simulation set only.
averaging over all realizations. The solid lines are for quasar lifetime
$t_{q}=10^{7}$ y and the dashed lines for $t_{q}=10^{8}$ y.
The solid points are the observational data computed from the Sloan Digital 
Sky Survey First Data Release.
}
\end{figure}

One way of reducing the relative effect of the radiation
from an individual quasar is by increasing the strength of the background
field. Of course, for the regular proximity effect, this has long been used
as a measure of the background field itself, given
a known flux of radiation from the quasar. Because of these
measurementsscott and other considerations, the amount of background radiation
not coming from QSOs is not a free parameter, but is constrained to 
be about equal to that from the sum total of QSO radiation (e.g., Sokasian
\etal 2003). The observational errors are large (e.g., for Scott \etal 2000,
the allowed intensity can vary by $+50\%, -60\%$ within $1 \sigma$).
 Nevertheless, it is instructive to vary 
the intensity of our additional homogeneous radiation field much more 
than this in order to investigate how much it can change
the foreground proximity effect. In Figure \ref{favuniform}, we 
show the $\langle F\rangle(r)$ which results when the background field is
changed from our fiducial value (equal to the QSO
contribution) to 64 times that value. As expected, increasing 
the background reduces the effect of the local radiation. However, even
with a background 16 times too high, there is still some
evidence of a proximity effect, and the simulations are far from the
observational points. Even if QSOs are responsible for a tiny fraction
of the intensity of the ionzing radiation field at $z=3$ (which 
does not seem likely), this will not solve our discrepancy.

\begin{figure}[t]
\centering
\psfig{file=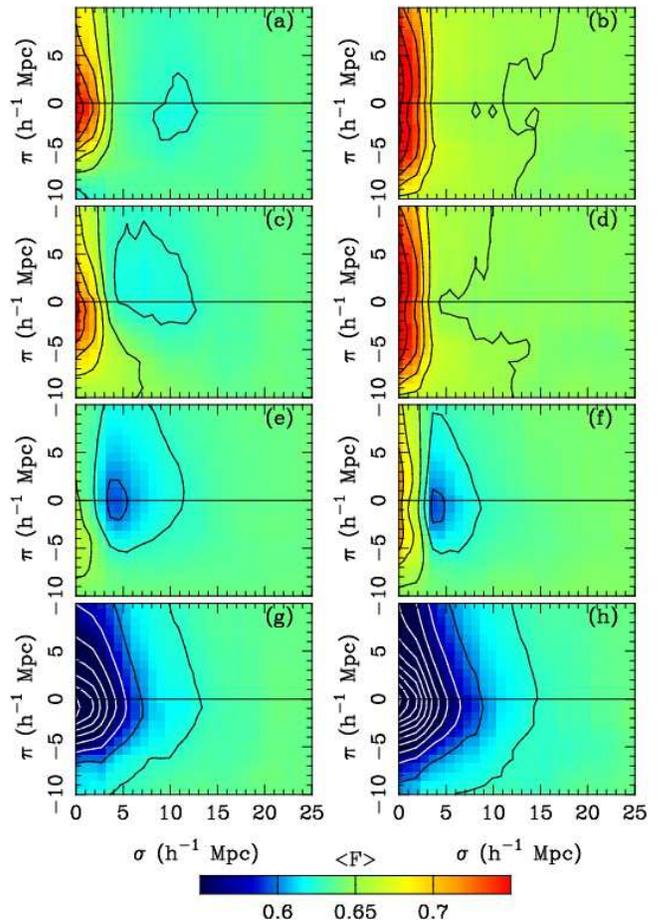,angle=-90.,width=8.5truecm}
\caption[]{ 
The mean \lya\ forest flux $\langle F \rangle$
as a function of distance across ($\sigma$)
and along ($\pi$) the line
of sight from simmulated quasars.
The left panels are for quasars with a lifetime
$t_{q}=10^{7}$ yrs and the right panels for 
$t_{q}=10^{8}$ yrs. Panels (a) and (b) are not output on the lightcone
and have no quasar beaming. Panels (c) and (d) include lightcone effects
(the observer's line of sight is the y-axis)  and 
isotropically emitted radiation. In panels (e) and (f), radiation
is emitted from quasars  in a cone of opening angle 90 degrees, and 
lightcone effects are included (see text). In panels (g) and
(h) the radiation field is uniform.}
\label{sigpisim}
\end{figure}

Additional information is available to us if we turn to the
plot of $\langle F\rangle$ in $\sigma, \pi$ space. We show this for the
simulations in Figure \ref{sigpisim}. As with the panels of Figure \ref{slices}
we how results for the two $t_{q}$ values and for no lightcone,
light cone effects and lightcone plus beaming.  The bottom panel 
shows the result for a uniform radiation background.
The top and bottom panels should both have symmetries about the $\pi=0$
axis, and any deviation from this gives a rough indication of the 
level of noise in the theoretical predictions.

The top panels show that the proximity effect region is stretched
appreciably along the line of sight by redshift space distortions.
The suppression of absorption caused by the rarer quasars with $t_{q}=10^{8}$
years is evident also. Moving down to the panels where lightcone
effects are included, the results for  $t_{q}=10^{8}$ are not 
noticeably different to the no lightcone plot. The shorter lived quasars
do show evidence of the expected asymmetry about the $\pi=0$ axis, however,
with a greater proximity effect (less absorption) towards negative
values of $\pi$. The region with above average absorption (a dark green 
area) can also be seen to lie more towards +ve $\pi$ values,
effectively shielded from some of the ionizing radiation by the
finite travel time of light in that direction. If this asymmetry were 
observable in data, it might be possible to use the information
contained in it to constrain cosmology. For example, the size of the
asymmetry depends both on the QSO lifetime and the physical 
scale associated with an angular and redshift scale.

In the third row, we have added the effect of anisotropically 
emitted radiation. Again an asymmetry about $\pi=0$ is visible as we
are including lightcone effects. The most striking difference between these
and the panels directly above is the large area which is protected 
from ionizing radiation. In these regions, which are off axis 
with respect to the line of sight, we expect the \lya\ forest flux to 
be similar to that for the simulation with a uniform radiation 
background.  Plotting the results as a function of $\sigma$ and $\pi$
allows us to see this structure which is not visible in Figure \ref{fav}
for example, and allows us to see the effect of beaming and 
constrain the opening angle directly. From this plot, even though all quasars
have been given the same beaming angle, one would not
expect to see a sharp shadow edge, because the center of the emitted radiation
cone is not always pointing along the line-of-sight. Also, because of this,
even regions which are close to the sightline can be outside the radiation
cone sometimes, so that the proximity effect is not as strong as for the 
panels with no beaming. Comparing to the SDSS results in 
Figure \ref{sigpisdss} it does not seem that the lack of a proximity
effect in that data can be due to the beaming from an angle
close to that we have used here (90 degree cone angle). Reproducing
Figure \ref{sigpisdss} with beaming alone would require a much smaller
angle. We must be cautious on two counts, however, as Figure \ref{sigpisdss}
was smoothed with a filter, so that absorption close to the $\pi$-axis 
can contain information from pixels farther away. This is particularly true
as there are very few quasar pairs with small angular separations, only 
5 quasars with sightlines
that come withing $5 \hmpc$ of another in the transverse direction exist.
Schirber et al (2003) present a constraint on the beaming angle from
a combination of numerical and analytic modelling. In our case, 
changing the beaming angle would change the nature of the QSOs and their 
host galaxies, as with a smaller angle they would need to be more common
(assuming only one quasar phase per host galaxy). They would
exist in different environments, which would also change the
absorption profile.

Finally, the bottom two panels of  Figure \ref{sigpisim} show the
$\langle F\rangle (\sigma,\pi)$ results for a uniform radiation 
field. The results 
must be isotropic with respect to changes in angle, so that small asymmetry
present about $\pi=0$ is again due to cosmic variance. The plot has a similar
appearance to Figure \ref{sigpisdss}, although there is not quite as much 
absorption close to QSOs,  indicative that real QSOs may be found even denser
environments than in the simulation.

\section{Summary and discussion}

\subsection{Summary}

In this paper, we have carried out a study of the
inhomogeneous large scale radiation field at redshift $z=3$ and its
effect on the \lya\ forest. We have combined dark matter and hydrodynamic
simulations to produced a prediction for this radiation field, in
boxes of size up to $500 \hmpc$ on a side. Using several different
box sizes and particle masses has enabled us to check convergence
of our results. Using these
simulated radiation fields, we have found the following in the
main part of the paper:  

(1) The structure in the radiation field is rich and complex,
with inhomogeneities most obvious on large scales, close to 
the mean path length of photons ($\sim 100 \hmpc$). The finite light 
travel time across these regions means that quasar light 
echos are an obvious feature of maps of the radiation field,
with output on the light cone leading  to crescent shaped
features.

(2) Averaged in cells of width 3 $\hmpc$, the radiation field
has RMS variations of between 0.58 and 1.1 times the 
mean, depending on whether 
sources have long lifetimes or not.
Shadowing of an individual source by neighboring filaments can cause
the radiation intensity due to the source
at distances of a few 10s of Mpc to vary by $50\%$.

(3) The effect of the inhomogeneous radiation
field on the average clustering properties of the 
\lya\ forest is relatively subtle. For both quasar lifetimes
we have tried, $10^{7}$y and $10^{8}$ y, the power spectrum of the flux
is suppressed by about $10\%$ on scales of $k =0.05-0.5 \invhmpc$.
We find that shadowing effects are not very important, and using isotropic
attentuation around each source only changes the prediction for
the suppression of P(k) by  $20 \%$.

(4) With both quasar lifetimes,
we predict that a large foreground proximity effect should
be seen in the  \lya\ forest of spectra that  pass close to 
other quasars.  The \lya\ forest transmitted flux  when
averaged around foreground quasars is predicted to have an upturn on 
scales $r \simlt 2 \hmpc$ and be significantly different from
the homogeneous  radiation field case for $r \simlt 10 \hmpc$.

(5) With a uniform radiation field (which would result if quasars
only accounted for  $<< 0.1$ of the
total intensity at 1 Ryd, which is unlikely), we predict significantly
more absorption around quasars, with the transmitted flux being less
than 0.5 on scales $r < 4 \hmpc$.

(6) We have used data (1920 quasar spectra)
 from the Sloan Digital Sky Survey First Data Release 
(Abazajian \etal 2003)
to carry out a measurement of the \lya\ forest transmitted flux in pixels 
averaged around foreground quasars. We find no evidence of 
a foreground proximity effect, but instead increased absorption close
to quasars, with transmitted flux being $F<0.5$ for $r < 7 \hmpc$.
This strong absorption is even more than that expected in the simulation
for the case with a uniform radiation field.

\subsection{Discussion}

The \lya\ forest is starting to become a useful cosmological tool, 
and inferring clustering properties of the mass
distribution from those of the \lya\ forest flux has been used to put 
constraints on cosmological models. Before clustering had been measured
in the distribution of the \lya\ forest (e.g., Webb 1987),
it was expected that radiation fluctuations would cause clustering 
themselves. It is obviously necessary to explore this, and include as many
of the relevant effects as possible. For example, shadowing,
the discreteness and clustering of sources and their possible
beaming, all combine to make structure in the radiation field potentially
very complex. In this paper, we have simulated the radiation field in a 
specific model, one in which the mean  intensity is mostly contributed
by quasars with relatively long lifetimes. In the context of this model,
we have made detailed predictions for the clustering of the \lya\ forest. 
In particular we have found that on large scales that are just beginning to 
be accessed by todays large quasar surveys, the power spectrum
of the flux is suppressed. This effect  could change the amplitude 
of matter clustering, $\sigma_{8}$ by as much as $\sim 7\%$?
(\S 4.2). As the radiation field on smaller scales is smoother, so that 
there is no suppression,  this
will tend to make the P(k) bluer. For example the P(k) suppression
seen over the range $k=0.1-1 \invhmpc$ in 
Figure \ref{conv} could change the inferred power law index $n$ 
of P(k) by $\sim +0.05$. This is assuming that the statistical 
weight in a determination
of the slope comes equally from each interval in $\log k$. In practice,
 this is not the case because of the large number of modes contributing
at small $k$ leads to smaller error bars so that the effect will be less.
This effect could still perhaps masquerade as a rolling spectral
index (non-zero value of ${\rm d}\log{n}/{\rm d}k$), although in the opposite
direction to that hinted at in the WMAP data papers (Spergel \etal 2003).

Because we have simulated perhaps the most extreme model for the radiation 
background (rare long lived quasar sources), this suppression of the 
\lya\ forest power spectrum is unlikely to be exceeded in other models.
However, at higher redshifts, the attentuation length will be much 
shorter, and the effect on the flux power spectrum much greater. Meiksin
\& White (2003b) have suggested that this statistic could actually be
used to constrain the ionizing background intensity field rather than the
clustering of matter at redshifts $z> \simgt 5.5$. For example they find 
that the flux $P(k)$ maybe even show an upturn on large scales 
$k < 0.5 \invhmpc $
for high enough redshifts ($z \sim 6$).
 In this paper, we have seen that perhaps rather 
surprisingly the effect of finite quasar lifetimes (not included
by Mekisin \& White, who used much smaller simulation volumes)
whilst complicating
the radiation field visually has little effect on the flux power spectrum.
It is possible, however that the higher order statistics of the
flux (see e.g., Gazta\~naga \& Croft 1999, Mandelbaum \etal 2003) are
affected. In future work, it will be interesting to measure
the bispectrum of the flux, which Mandelbaum \etal (2003) have shown can
potentially be used to check on the gravitational nature of the mechanism
causing growth of  clustering. Viel \etal (2003) have measured
the bispectrum of the flux for a sample of 27 high resolution,
high signal to noise spectra, and find results consistent with simulations
which assume a uniform radiation background, and with an analytical model for
weakly nonlinear gravitational instability. These results are
however on smaller scales than we are able to simulate here.

In this paper we have not made a comparison of our flux P(k) results with 
observational data on large scales. This data will be forthcoming, 
for example from the SDSS. We have however, used the Sloan data to try
to make a measurement of the foreground proximity effect, and found no evidence
for the effect of quasar radiation on \lya\ spectra in sightlines passing 
nearby. This was very different to what was predicted from the simulations.
Two questions then present themselves. First, in what way could
the real Universe be different from our models which could
account for this difference. Second, what are the
implications for our predictions
of the effect of radiation fluctuations on P(k)?

As mentioned previously, Schirber and Miralda-Escud\'{e} (2003) found the
lack of a foreground proximity effect when looking at SDSS \lya\ 
forest spectra for which the sightlines are close to  three specific
bright quasars. They were chosen because based on 
quasar luminosites, their radiation should have overwhelmed the 
background level by factors of 13-94.  These authors investigated 
in detail several
possible reasons why no foreground proximity effect was seen. These
were: gas density greater than the cosmic mean close to quasars, 
anisotropic radiation emission, and a short  quasar lifetime
($t_{q} < 10^{6}$ years). They found that 
each one of these effects was unlikely to be responsible,
but that a combination of all three was not unreasonable. 

In this paper, we
have used a much larger sample of quasars, but the number of quasars
which are close to another sightline is still small. Also, we
are averaging over all quasars, even those for which the locally
produced radiation does not overwhelm the background by a large factor 
(there are not any more of these overwhelmingly bright quasars in our 
sample, Schirber 2003, {\it private communication}).
As our simulations do place quasars in overdense regions, we 
can say in the context of our model that this does not
cancel out the expected foreground proximity effect. Of course, the
fact that a line of sight proximity effect has perhaps been seen out distances
of $> 40 \hmpc$ (comoving) (e.g., Dobrzycki \& Bechtold 1991,
and Jakobsen \etal 2003 for HeII) mean that this is not
likely to be the whole story in any case.

The anisotropy of quasar emission was investigated in our simulations
using a half-opening angle of 45 degrees. This did have a noticeable
effect on the absorption plotted in the $\sigma -\pi$ plane, with shadowing
evident of regions at greater angles from the sightline. The extra
absorption in these regions did not lead to much difference in the 
angle-averaged mean absorption around quasars though, and in order
to reproduce the observed results, a very small opening angle would
appear to be required. For example, in the observational
sample, there are 5 sightlines
with an impact parameters between $2.5-5 \hmpc$, and these show no 
evidence of a proximity effect. The excess absorption over the mean 
seen close to quasars is evident out to $10 \hmpc$, which means that 
a maximum half opening angle of $\sim 15\deg$ is required. As also calculated
by Schirber and Miralda-Escud\'{e} (2003), this seems too small
to be consistent with expectations of quasar emission.
Another argument against a very small  angle is that 
both the clustering of quasars and the absorption around them would
likely be much reduced, as the actual space density of quasars would
be much higher, and we would sampling lower mass halos. The correlation
function of our quasars ( $r_{0}$ =6.5 for $t_{q}=10^{7}$y) is already
consistent with that of the 2dF quasar survey with either isotropic
emission or our $45 \deg$ opening angle, and the excess \lya\ absorption
seen close to quasars is already stronger in the SDSS data  than for our
 most massive simulated quasars
(we shall return to this below). 

The lifetime of quasars will also have an effect on whether a
foreground proximity effect is seen. In this paper, we have concentrated
on relatively long-lived quasars, with $t_{q}=10^{7}$ or $t_{q}=10^{8}$.
There are a number of arguments which suggest that the averaged lifetime
should be within this range. For example, we have already mentioned the
clustering strength of QSOs (e.g., Martini \& Weinberg 2001, Haiman and 
Hui 2001) which seems to require $t_{q} > 10^{6}$y. 
There are also arguments based
on the fraction of Lyman-break galaxies which have AGN (Steidel \etal 2001),
which also point to values of $t_{q}\sim 10^{7}y$. Theoretically,
it is possible to make models of the AGN population
and predict observables. These models often 
set $t_{q}$ to be a free parameter. An approch which uses a hydrodynamic
cosmological simulation of galaxy formation as a starting point 
to do this was presented in Di Matteo \etal (2003). Consistency
with the observed luminosity functions and mass density in blackholes
was found with $t_{q} \sim 2\times 10^{7}$y. An alternative, model-based
approach to constraining $t_{q}$ involves comparing the 
metallicities of gas in the AGN broad line regions (e.g., from
the observations by Dietrich \etal 2003) with model predictions.
For longer $t_{q}$, quasars are more massive objects and tend to have 
have higher circumnuclear metallicities. Again, values of 
$t_{q}=10^{7}-4\times 10^{7}$ years
are consistent with current observations (Di Matteo \etal 2003b).

As emphasized by Schirber \etal (2003),
 all of the above methods constrain the total time
of emission of each quasar. This is different to the length in time of 
the last burst of ionizing radiation, which is what is relevant
with the foreground proximity effect. If quasars vary on relatively short
timescales, so that this total time of emission is composed of
short bursts, then they could still
potentially be associated with massive halos 
and inhabit dense environments, but could avoid our bounds on the
proxmity effect. A lower bound on the length of bursts in this
variable scenario is given by consideration of the line of sight
proximity effect. The photoionization timescale relevant for the 
the IGM to respond to an increase in the intensity of ionizing radiation
 is approximately $10^4$ yrs,
so that quasars must be on for at least this amount of time. Another
lower limit on the length of burst is set by observing large numbers
of quasars at different epochs in order to see 
directly if any have switched off or
on, or dimmed or brightened. With enough quasars and a long 
enough time between epochs, this can result in a limit which is
competitive (for example Martini \& Schneider 2003 found $t_{q}> 2\times10^{4}$
from analysis of the
SDSS Early Data Release). These lower bounds are fairly
short compared to the arguments related to the total time of emission.
However, there are some cases where it seems
as though the quasars must have been shining continously for a much 
longer period of time. These include the \lya\ forest void
possibly caused by a foreground quasar seen by Dobrzycki \& Bechtold (1991).
Schirber \etal (2003) also point out another argument based on the large
size of the line-of-sight proximity effect regions 
in the HeII forest seen around 
 PKS1935 (Anderson \etal 1999) and
Q0302-003 (Heap \etal 2000). They argue that the
regions are so large ($\delta_{z}\sim 0.08$) that shining at their presently
observed luminosities these quasars would have had to do so for $\sim 10^{7}$
years in order to maintain the level of ionization seen. 

In our case, we see no evidence for a foreground proximity effect,
which would seem to necessitate the time since the last burst being
substantially shorter than the
shortest lifetime in our simulations, $t_{q}=10^{7}$ y. We have however already
seen that there are reasons, including the large amount of 
absorption close to quasars to assume that the total time
of emission is at least this value, and probably more.
A scenario whereby AGN are strongly variable on timescales
of $10^4 y < t_q < 10^{6}$y  but each quasar has a total time
of emission greater than $t_{q}=10^{7}$ might perhaps be a solution (except
for the arguments at the end of the previous paragraph).

In our case, we would need the average time between bursts for
each quasar to be much longer than the length of bursts, so that  the
foreground proximity effect from the previous burst would not be
easily seen. In order to put proper constraints on the length of bursts,
we should run a self consistent model  
with short separate bursts orginating from the same quasar at widely
spaced intervals. As we see no evidence for a foreground 
proximity effect from quasars closest to the line of sight ($2.5 \hmpc$),
it is likely that the length of bursts should be less than the 
light travel time (3 Myr). In such a scenario, the radiation field 
might be more patchy, because of the smaller thickness
of the light echoes, but then again there would be many more of them.
 In Figure \ref{fpkratio} we have seen that the 
 changing $t_{q}$ from $10^{8}$ y to $10^{7}$ y  does little to
the suppression of large scale power in the flux power spectrum,
so that adopting short bursts of $10^{6}$ might  not change this much.
However, on small scales, for $10^{7}$ y, the power spectrum has
not quite converged, so it is possible that the smaller scale
power would be affected by a $10^{6}$ y burst lifetime. This possibility,
shorter bursts, should be investigated directly using smaller simulations
at higher resolution, and we plan to do this in the future.

 Although we have seen that there is 
some evidence that quasars do shine for longer continuous periods, the
short bursts scenario could perhaps help with some observations.
For example, Shull \etal (2003) have found that the ratios of column densities
of HI and HeII absorbers vary from place to place, with variations
taking place over short scales, $\sim 3 \hmpc$ comoving. This
behaviour is consistent
with variations in the spectrum of ionizing radiation
on these scales. How this arises is another matter, as the mean
separation between bright quasars is of the order of 40 times larger.
 Shull \etal attribute these variations 
to the result of non-uniformities in opacity, coupled with the wide observed
range in QSO spectral indices. In this paper we have
only simulated the propagation of radiation with energy 1 Ryd, so that 
we cannot directly model the variation of the spectrum
of radiation which arises from the different propagation of harder radiation.
However, if quasars emit their ionizing radiation in short bursts 
of length $\sim 10^{6}$y, and there is a wide range of
quasar spectral indices, then fluctuations in the spectrum of
radiation on scales $\sim 1 \hmpc$ (the thickness of light echoes)
might be expected, even without large opacity variations.

Spatial fluctuations in the ionizing radiation field are not
the only effects which could cause fluctuations in the
\lya\ absorption. One effect which we have not simulated here
are variations in the temperature of the gas, specifically due
to late HeII reionization. There is some evidence that the reionization
of HeII took place at around $z=3$. For example, indirect 
evidence from the ratios of
CIV and SIV absorption lines, which are sensitive to 
the radiation spectrum
(e.g., Songaila \& Cowie 1996). Some more direct
information comes from the observed patchiness of HeII \lya\ forest absorption,
and the large clearings observed in HeII around quasars 
mentioned above (see also Heap \etal 2000). The local heating 
which occurs as HeI is ionized to HeII will change the 
temperature-density relation, making it vary spatially. This will change
the ionization balance of the gas which contributed to the \lya\ forest,
with increasing temperatures causing more collisional ionization and 
less absorption. This will be related spatially to the distribution of
sources (presumably quasars, because of the necessity for hard photons
of energy $> 4$ Ryd), and will cause spatial fluctuations in the forest.
It is beyond the scope of this paper to model these temperature fluctuations,
but we note that they will have a different character, as the timescale
to cool back to the mean $\rho-T$ relation is much longer 
(Hui \& Haiman 2003) , and so the sharpness of features seen 
in the maps of the radiation field (e.g. Figure \ref{slices})
will not be present.
Detecting these fluctuations may also be challenging. We note that 
observationally in this paper
 we have seen only increased absorption around the SDSS
quasars, so that neither the radiation proximity effect nor
the effect of HeII reionization temperature increases has been evident.
Globally, the signature of a rise in temperature
and isothermality of the equation of state at $z\sim3$ has been
seen by Shaye \etal 2000 and Ricotti \etal 2000
  (although see McDonald \etal 2001).
A small decrease in the mean absorption 
(a feature in the \lya\ forest optical depth vs redshift)
when averaged over all \lya\ forest pixels 
at this redhift has also been seen by Bernardi \etal (2002). Bernardi
\etal interpreted this to be the signature of HeII reionization (see
Theuns \etal 2002 for theoretical modelling).

There are many opportunities for future work to
investigate the fluctuations
in the radiation field.  As the density field has a structure which 
was dramatically revealed on large scales by the first galaxy redshift
surveys, the ionizing radiation field may also vary from place to place
in a manner which is different but equally complex. The same 
simulation techniques used here could be applied to higher redshifts,
where the attenuation length becomes much smaller and the radiation
induced fluctuations in the \lya\ forest much larger. Larger and 
higher resolution hydrodynamic simulations would eliminate the 
need to combine dark matter and hydro runs together in the way we
have (such runs are already becoming available, Springel \& Hernquist 2003,
{\it private communication}).
 The simple radiative transfer techniques we have used
could also be replaced with time dependent codes, and 
more than one frequency of radiation could be followed. As mentioned in
the introduction, many papers have been written about the reionization
epoch by groups working with these approaches. We plan to improve our
technique with an iterative scheme to model the effect of local
ionizing radiation from a source on propagation of its own radiation.
The observational dataset we have used will increase by a factor 
of $\sim5$ as the Sloan Digital Sky Syrvey nears completion, and better
constraints on the nature of sources and  quasar lifetimes  will
become possible. As the nature of the ionizing radiation field becomes
better understood, it may even become possible to
use such observations to constrain cosmology (see e.g., Phillips \etal 2002).

\bigskip
\acknowledgments 
We thank Lars Hernquist for the use of facilities at 
the Harvard-CfA Center for Parallel
Astrophysical Computing. Some simulations were also carried out on 
the CMU Astrophysics parallel cluster.
We thank Lars Hernquist and Volker Springel for providing a hydro simulation 
output and George Efstathiou for the use of the $P^{3}M$ code.
We also thank Chris Miller for help with the Sloan 
Data and Lars Hernquist, Tom Abel, Michael Schirber, Tiziana Di Matteo,
Simon White, Rashid Sunyaev, and Avery Meiksin for useful discussions,
and the Max Planck Institute
for Astrophysics for hospitality during visits.

\end{document}